\documentclass{aa}
\usepackage{natbib}
\usepackage{graphicx}
\usepackage{bm}
\usepackage{hyperref}
\usepackage[capitalise]{cleveref}
\usepackage{txfonts}
\bibpunct{(}{)}{;}{a}{}{,}
\usepackage{braket}

\hypersetup{
    colorlinks=true,
    citecolor=blue,
    linkcolor=blue,
    breaklinks=true,
}

\creflabelformat{equation}{(#2\textup{#1}#3)}
\creflabelformat{figure}{(#2\textup{#1}#3)}

\renewcommand*\d{\mathop{}\!\mathrm{d}}
\newcommand{\ii}{\mathrm{j}}
\newcommand{\bigbraket}[2]{\Braket{\vphantom{a_a^a}#1|#2}} 
\newcommand{\po}{\Psi_\text{osc}}
\newcommand{\pt}{\Psi_\text{turb}}
\newcommand{\fpo}{\widehat{\Psi}_\text{osc}}

\newcommand{\uxo}{\widehat{u}_{x,\text{osc}}}
\newcommand{\uxt}{\widehat{u}_{x,\text{turb}}}
\newcommand{\uyo}{\widehat{u}_{y,\text{osc}}}
\newcommand{\uyt}{\widehat{u}_{y,\text{turb}}}
\newcommand{\vo}{\widehat{\zeta}_\text{osc}}
\newcommand{\vt}{\widehat{\zeta}_\text{turb}}
\newcommand{\ffpt}{\widetilde{\Psi}_\text{turb}}
\newcommand{\nut}{\nu_\text{turb}}
\newcommand{\ret}{\mathrm{Re}_\mathrm{turb}}
\newcommand{\yo}{y_\text{obs}}

\newcommand{\ep}{\mathcal{E}_\Psi}
\newcommand{\eu}{\phi_{ij}}
\newcommand{\oeq}{\Omega_\mathrm{eq}}
\newcommand{\eiso}{E_\mathrm{iso}}

\begin{document}

\title{Interaction of solar inertial modes with turbulent convection}
\subtitle{A 2D model for the excitation of linearly stable modes}
\titlerunning{Stochastic excitation of solar inertial modes}
\author{J. Philidet\inst{\ref{inst1}} \and L. Gizon\inst{\ref{inst1},\ref{inst2}}}
\institute{Max-Planck-Institut f\"ur Sonnensystemforschung, Justus-von-Liebig-Weg 3, 37077 G\"ottingen, Germany \label{inst1}
\and
Institut für Astrophysik, Georg-August-Universit\"at G\"ottingen,  37077 G\"ottingen, Germany \label{inst2}
}
\date{Received 12 December 2022 / Accepted 28 March 2023}

\abstract
{Inertial modes have been observed on the Sun at low longitudinal wavenumbers. These modes probe the dynamics and structure of the solar convective zone down to the tachocline. While linear analysis allows the complex eigenfrequencies and eigenfunctions of these modes to be computed, it gives no information about their excitation nor about their amplitudes.}
{We tested the hypothesis that solar inertial modes are stochastically excited by the turbulent motions entailed by convection. Unlike the acoustic modes, which are excited by vertical turbulent motions, the inertial modes are excited by the radial vorticity of the turbulent field.}
{We have developed a theoretical formalism where the turbulent velocity fluctuations provide the mechanical work necessary to excite the modes. The modes are described by means of a 2D linear wave equation with a source term, under the $\beta$ plane approximation. This wave equation restrained to a spherical surface is relevant for the quasi-toroidal inertial modes that are observed on the Sun. Latitudinal differential rotation is included in the form of a parabolic profile that approximates the solar differential rotation at low and mid latitudes. The turbulent vorticity field underlying the source term is treated as an input to the model and is constrained by observations of the solar surface. The solution to the linear inhomogeneous wave equation is written in terms of a Green function, which is computed numerically.}
{We obtain synthetic power spectra for the wave's latitudinal velocity, longitudinal velocity, and radial  vorticity, with azimuthal orders between $1$ and $20$. The synthetic power spectra contain the classical equatorial Rossby modes, as well as a rich spectrum of additional modes. The mode amplitudes are found to be of the same order of magnitude as observed on the Sun ($\sim 1$~m/s). There is a qualitative transition between low and high azimuthal orders: the power spectra for $m \lesssim 5$ show modes that are clearly resolved in frequency space, while the power spectra for $m \gtrsim 5$ display regions of excess power that consist of many overlapping modes.}
{The general agreement between the predicted and observed inertial mode amplitudes supports the assumption of stochastic excitation by turbulent convection. Our work shows that the power spectra are not easily separable into individual modes, thus complicating the interpretation of the observations.}


\maketitle

%%% Introduction

\section{Introduction}

Multiple types of waves can propagate in the interior of a star. In the case of a non-rotating star, these modes are spheroidal; they are the $p$-modes (or acoustic modes), the $f$-modes (or surface-gravity modes), and the $g$-modes (or gravity modes). The first two have been observed on the Sun for a long time \citep{leighton62,deubner75} and are used to probe the equilibrium structure of the solar interior \citep[see][for a review]{jcd02}. Gravity modes, by contrast, are evanescent throughout the solar convective envelope, precluding us from using them as probes of the solar radiative interior.

The Sun, however, is a rotating star, and the inclusion of rotation entails the possibility of additional modes of oscillation. In a uniformly rotating star, theory predicts the existence of quasi-toroidal modes, known as $r$-modes \citep{papaloizou78}, with the Coriolis force as the restoring force. In the rotating frame, they propagate in the retrograde direction and have frequencies comparable to the rotation rate. They are similar to the Rossby waves that are ubiquitous in the atmosphere of the Earth \citep{rossby39} and other planets \citep[e.g.][]{allison90,sanchez-lavega14}. Equatorial Rossby modes were recently observed on the Sun \citep{loptien18,liang19} with a dispersion relation close to that of the theoretical sectoral ($l = m$) modes. A much richer spectrum of modes, collectively referred to as inertial modes, was subsequently reported by \citet{gizon21}; in addition to the equatorial Rossby modes, these include high-latitude modes and critical latitude modes, allowed by latitudinal differential rotation. Furthermore, \citet{hanson22} report the observation of high-frequency waves of vorticity that are anti-symmetric with respect to the equator.

These inertial modes allow the convective zone  to be probed in a way that complements p-mode helioseismology, especially when it comes to the superadiabatic temperature gradient and the turbulent viscosity. To do so, it is necessary to develop a theoretical understanding of the physics of inertial modes. \citet{gizon20} carried out a linear analysis of viscous modes of a parabolic shear flow in the $\beta$ plane. This analysis, which applies to toroidal modes, was subsequently extended to viscous modes on a sphere by \citet{fournier22}, thus allowing for a more realistic differential rotation profile and a treatment of the lowest azimuthal orders. A linear analysis of the 3D solar convection zone was carried out by \citet{bekki22}, who identified several of the modes reported by \citet{gizon21}. Likewise, \citet{triana22} proposed an identification of the modes reported by \citet{hanson22}.

Such linear analyses give information about the frequencies, the stability, and the eigenfunctions of the inertial modes, and allow for the identification of some of them in the observations. They also reveal that some of these modes can be linearly unstable, as a result of strong latitudinal differential rotation \citep{fournier22} or a baroclinic instability due to a latitudinal entropy gradient \citep{bekki22}. However, for latitudinal differential rotation profiles that are not too pronounced, most of the modes are predicted to be linearly stable, meaning that they are likely excited by turbulent convection. Understanding the excitation process of the linearly stable solar inertial modes would not only allow us to  put stronger constraints on the dynamics of the solar convective zone, but would also help us predict which modes are expected to be visible and identifiable. This would go a long way towards helping us interpret the observational data at our disposal. In the present paper, we do not address the case of unstable inertial modes.

In the absence of a destabilising mechanism, the turbulent motions associated with solar convection provide an excitation mechanism, as in the case of solar $p$-modes \citep{lighthill67,goldreich77b} or gravito-inertial waves \citep{mathis14,augustson20}. Non-linear 3D simulations of the convection zone can help us assess the importance of this mechanism \citep{bekki22b,dikpati22}. In this paper, we follow a different approach and present a theoretical model for the turbulent stochastic excitation of purely toroidal inertial modes in 2D in order to test the hypothesis that this mechanism is indeed responsible for the amplitude level at which inertial modes are observed on the Sun. Because the analysis is done in 2D, it is only relevant for the predominantly toroidal modes, such as the equatorial Rossby modes and the other inertial modes that have been observed. Furthermore, we place ourselves in the equatorial $\beta$-plane approximation, similarly to \citet{gizon20}. We assume that the inertial modes are excited by turbulent emission, meaning that the non-linear advection term in the momentum equation plays the role of a source term in the linear wave equation. This is also in accordance with the commonly accepted picture for $p$-modes \citep[e.g.][]{samadi01}.

The paper is organised as follows. We present the stochastic excitation formalism in Sect. \ref{sec:SyntheticSpectrumModel}. The formalism requires two main ingredients, namely the Green function associated with the homogeneous wave equation (which is computed numerically, as described in \cref{subsec:GreensFunction}), and the convective turbulent spectrum underlying the source term (which is treated as an input to the model and is the subject of Sect. \ref{subsec:TurbulentSpectrum}). This model provides us with synthetic power spectra containing both the normal inertial modes of the system and the turbulent noise responsible for their generation. We focus on the expectation value of the power spectrum near the solar equator in Sect. \ref{subsec:EquatorialSpectra} and discuss the latitude dependence of the power spectra in Sect. \ref{subsec:2DSpectra}. Conclusions are drawn in Sect. \ref{sec:conclusion}.

%%% Section 2

\section{Synthetic power spectra\label{sec:SyntheticSpectrumModel}}

\subsection{Stochastic excitation by turbulence\label{subsec:AnalyticalModel}}

We study the excitation of the vorticity modes observed on the Sun. These modes are quasi-toroidal (characterised by their horizontal motions); we assume that the horizontal part of the wave equation can be decoupled from the radial part. In the following, we focus on the horizontal part and study the excitation of vorticity waves in a 2D shear flow mimicking the solar differential rotation \citep{gizon20}. We place ourselves in the equatorial $\beta$ plane, where the 2D spherical coordinates $\lambda$ and $\phi$ (denoting respectively the latitude and longitude) are transformed into
\begin{align}
    & x \equiv R\phi~, \\
    & y \equiv R\sin\lambda~,
\end{align}
where $R$ is the radius of the Sun. The velocity components on the sphere ($\mathrm{v}_\lambda$ and $\mathrm{v}_\phi$) can be approximated by the Cartesian components in the $\beta$ plane ($v_x$ and $v_y$). The equations of motion in the rotating frame and in the inviscid limit become
\begin{align}
    & \dfrac{\partial v_x}{\partial t} + (\mathbf{v} \cdot \bm{\nabla}) v_x = -\dfrac{1}{\rho} \dfrac{\partial p}{\partial x} + f v_y~, \label{eq:BetaPlaneFlowX} \\
    & \dfrac{\partial v_y}{\partial t} + (\mathbf{v} \cdot \bm{\nabla}) v_y = -\dfrac{1}{\rho} \dfrac{\partial p}{\partial y} - f v_x~,
    \label{eq:BetaPlaneFlowY}
\end{align}
where $\rho$ is the density, $p$ is the gas pressure, $f = \beta y = (2\oeq/R) y $ is the Coriolis parameter, and $\oeq$ is the rotation rate at the equator. A linear inhomogeneous wave equation can be derived from \cref{eq:BetaPlaneFlowX,eq:BetaPlaneFlowY}; the details are provided in \cref{app:WaveEquation}. To do so, the total flow velocity $\mathbf{v}$ is decomposed into a background zonal flow $\mathbf{U} \equiv U(y) ~ \mathbf{e_x}$, which represents differential rotation, and a residual flow $\mathbf{u}$, which contains both the waves and the turbulent noise. Working under the assumption that the residual flow is incompressible, we introduce the stream function $\Psi$ such that
\begin{equation}
    \mathbf{u} = \bm{\nabla} \wedge \left( \Psi \mathbf{e_z} \right)~,
\end{equation}
where $\mathbf{e_z}$ is the unit vector normal to the surface. This stream function is then decomposed into a contribution $\po$ for the oscillations and a contribution $\pt$ for the convective turbulent noise. Linearising in terms of $\po$ while keeping all orders in $\pt$, we obtain (see \cref{eqapp:WaveEquation})
\begin{multline}
    \left( \dfrac{\partial}{\partial t} + U\dfrac{\partial}{\partial x}\right) \Delta\po + (\beta - U'') \dfrac{\partial \po}{\partial x} - \nut \Delta^2 \po \\
    = \delta\left( \dfrac{\partial\pt}{\partial x}\dfrac{\partial \Delta\pt}{\partial y} - \dfrac{\partial\pt}{\partial y}\dfrac{\partial \Delta\pt}{\partial x} \right) ,
    \label{eq:WaveEquation}
\end{multline}
where $U''$ is the second derivative of $U(y)$, $\Delta = \partial_x^2 + \partial_y^2$ is the Laplacian operator, $\nut$ is the turbulent viscosity, and the operator $\delta$ denotes a fluctuation around the horizontal average taken on scales larger than the turbulence scale, but shorter than the wavelength of the inertial modes ($\delta q \equiv q - \langle q \rangle_h$ for any quantity $q$). The definition of this average requires a separation of scale, which is discussed in \cref{subsec:TurbulentSpectrum}. The left-hand side of \cref{eq:WaveEquation} governs the propagation of the inertial waves. As will be checked later, these modes are all linearly stable, because of the dissipative turbulent viscosity, $\nut$, which continuously pumps energy away from the modes, and because the shear induced by the differential rotation included in the model is not too strong. On the other hand, the right-hand side of \cref{eq:WaveEquation} acts as a source term that continuously injects energy into the modes. The equilibrium amplitude reached by the modes is the result of a balance between the damping and driving processes. Physically, the source term corresponds to the fluctuations of the divergence of the Reynolds stress tensor around its statistical average, and is stochastic by nature, because of the highly turbulent nature of the flow in the solar convective zone. The mechanism is similar to the traditionally accepted picture of solar-like p-mode excitation \citep{goldreich77b}; a key difference, however, is that whereas p-modes are mainly excited by the vertical turbulent motions, the quasi-toroidal inertial modes considered here are much more sensitive to the horizontal, vorticity component of turbulence.

The inhomogeneous wave equation can be written in terms of Fourier modes  $\exp(\ii\omega t - \ii k_x x)$, where $\ii$ denotes the imaginary unit. We used the following convention for the Fourier transform,
\begin{equation}
    \widehat{f}(\omega, k_x, y) \equiv \dfrac{1}{\sqrt{T_\mathrm{obs} X_\mathrm{obs}}} \displaystyle\int \d t \d x ~ f(t, x, y) e^{\ii(\omega t - k_x x)}~, \label{eq:DefinitionPartialFourier}
\end{equation}
where $T_\mathrm{obs}$ and $X_\mathrm{obs}$ are respectively the $t$ and $x$ windows over which the integral defining the Fourier transform is computed. In the Fourier domain, \cref{eq:WaveEquation} becomes
\begin{equation}
    \mathcal{L}\ \fpo = \widehat{S}~.
    \label{eq:WaveEquationFourier}
\end{equation}
The notation $\widehat{S}$ refers to the ($x$, $t$) Fourier transform of the right-hand side of \cref{eq:WaveEquation}, and $\mathcal{L}$ is the linear propagation operator, given by
\begin{equation}
    \mathcal{L}\equiv (\omega - k_x U) \widehat{\Delta} - k_x (\beta - U'') - \ii \nut \widehat{\Delta}^2~,
    \label{eq:DefinitionLinearOperator}
\end{equation}
where $\widehat{\Delta} \equiv \d^2 / \d y^2 - k_x^2$. The linear operator $\mathcal{L}$ and the source $\widehat{S}$ depend on $y$, the angular frequency $\omega$, and the longitudinal wavenumber $k_x$. 

The solution $\fpo$ to \cref{eq:WaveEquationFourier} can be expressed in terms of the Green function $G(y,y_s)$, which is defined as the solution of the following differential equation (and with the same boundary conditions as the full linear problem),
\begin{equation}
    \mathcal{L}\ G(y,y_s)  = \delta(y - y_s)~,
    \label{eq:DefinitionGreensFunction}
\end{equation}
where $\delta$ is the Dirac distribution, so that
\begin{equation}
    \fpo(y) = \displaystyle\int_{-R}^{R} \d y_s ~ G(y,y_s) ~ \widehat{S}(y_s)~.
    \label{eq:FormalSolutionStream}
\end{equation}
This solution leads to other physical quantities, such as the azimuthal velocity $\uxo$, the latitudinal velocity $\uyo$ and the radial vorticity $\vo$, for which we obtain
\begin{align}
    & \uxo(y) = \displaystyle\int_{-R}^R \d y_s ~ \dfrac{\partial G}{\partial y} ~ \widehat{S}(y_s)~, \label{eq:FormalSolutionUx} \\
    & \uyo(y) = -\ii k_x \displaystyle\int_{-R}^R \d y_s ~ G(y,y_s) ~ \widehat{S}(y_s)~, \label{eq:FormalSolutionUy} \\
    & \vo(y) = \displaystyle\int_{-R}^R \d y_s ~ \left(k_x^2 G - \dfrac{\partial^2 G}{\partial y^2} \right)~ \widehat{S}(y_s)~. \label{eq:FormalSolutionVort}
\end{align}
Then we obtain the expectation of the power spectral density by forming the modulus squared of \cref{eq:FormalSolutionUx,eq:FormalSolutionUy,eq:FormalSolutionVort} and taking the ensemble average. We assume that the spatial scale of the Green function and of the source are well separated -- the validity of this assumption will be checked in \cref{subsec:TurbulentSpectrum}. We find
\begin{align}
    & \left\langle \left| \uxo(\yo) \right|^2 \right\rangle = \displaystyle\int_{-R}^R \d y_s ~ \left| \dfrac{\partial G}{\partial \yo} \right|^2 \mathcal{I}(y_s)~, \label{eq:PowerSpectrumUx} \\
    & \left\langle \left| \uyo(\yo) \right|^2 \right\rangle = \displaystyle\int_{-R}^R \d y_s ~ \left| \vphantom{a^a_a} k_x G(\yo,y_s) \right|^2 \mathcal{I}(y_s)~, \label{eq:PowerSpectrumUy} \\
    & \left\langle \left| \vo(\yo) \right|^2 \right\rangle = \displaystyle\int_{-R}^R \d y_s ~ \left| \left(k_x^2 G - \dfrac{\partial^2 G}{\partial \yo^2} \right) \right|^2 \mathcal{I}(y_s)~, \label{eq:PowerSpectrumVort}
\end{align}
where $\yo$ is the latitudinal coordinate at which the power spectrum is evaluated, and $\langle ~.~ \rangle$ denotes an ensemble average. The function $\mathcal{I}(y_s)$ denotes the source covariance, and is defined by
\begin{equation}
    \mathcal{I}(y_s) \equiv \displaystyle\int \d Y ~ \left\langle \widehat{S}(y_s) ~ \widehat{S}^\ast(y_s + Y) \right\rangle~.
    \label{eq:DefinitionFastVariableIntegral}
\end{equation}
We recall that the source term $\widehat{S}$ depends on $\omega$ and $k_x$, so that $\mathcal{I}$ does too. The computation of the source covariance is detailed in \cref{app:SourceCorrelations}. We assume, in particular, that the source term is homogeneous, in the sense that its statistical properties do not depend on latitude; as such, the source autocorrelation spectrum no longer depends on $y_s$. Eventually, we find (see \cref{eqapp:FastVariableIntegral})\begin{multline}
    \mathcal{I} = \dfrac{1}{4\pi^3} \displaystyle\int \d\omega' \d^2\mathbf{k}' ~ k_x^3 k_x' k_y'^2 \left| \mathbf{k}' + \mathbf{k}/2 \right|^2 
    \\
    \times \ep\left(\omega' - \dfrac{\omega}{2}, \mathbf{k}' - \dfrac{\mathbf{k}}{2} \right) \ep^\ast\left(\omega' + \dfrac{\omega}{2}, \mathbf{k}' + \dfrac{\mathbf{k}}{2} \right)~.
    \label{eq:FastVariableIntegral}
\end{multline}
We note that the integrals span all frequencies and wavenumbers, positive and negative alike. The function $\ep$ represents the turbulent stream function spectrum, and is given by (see \cref{eqapp:DefinitionStreamFunctionSpectrum})
\begin{equation}
    \ep(\omega, \mathbf{k}) \equiv \displaystyle\int \d\tau \d^2\mathbf{x} \left\langle \vphantom{a_a^a} \pt(T, \mathbf{X}) \pt(T+\tau, \mathbf{X}+\mathbf{x}) \right\rangle ~ e^{\ii (\omega\tau - \mathbf{k} \cdot \mathbf{x})}~.
    \label{eq:DefinitionStreamFunctionSpectrum}
\end{equation}
We assumed that the turbulence is stationary and homogeneous, such that the turbulent spectrum depends on neither absolute time, $T$, nor on absolute space, $\mathbf{X}$.

\Cref{eq:PowerSpectrumUx,eq:PowerSpectrumUy,eq:PowerSpectrumVort} correspond to the contribution of the inertial modes to the velocity and vorticity power spectra. These spectra also contain a contribution from the turbulent noise, which can be expressed solely as a function of the turbulent stream function spectrum, $\ep$. Since this spectrum is the same quantity that appears in \cref{eq:FastVariableIntegral}, the contribution of the inertial modes and of the turbulent noise can be modelled simultaneously. We find
\begin{align}
    & \left\langle \left| \uxt(\yo) \right|^2 \right\rangle = \dfrac{1}{2\pi} \displaystyle\int \d k_y ~ \ep(\omega, \mathbf{k})~ k_y^2 ~, \label{eq:NoiseUx} \\
    & \left\langle \left| \uyt(\yo) \right|^2 \right\rangle = \dfrac{1}{2\pi} \displaystyle\int \d k_y ~ \ep(\omega, \mathbf{k})~ k_x^2 ~, \label{eq:NoiseUy} \\
    & \left\langle \left| \vt(\yo) \right|^2 \right\rangle = \dfrac{1}{2\pi} \displaystyle\int \d k_y ~ \ep(\omega, \mathbf{k})~ \left(k_x^2+k_y^2 \right)^2 ~. \label{eq:NoiseVort}
\end{align}
Forming the sum of the inertial mode contributions (i.e. \cref{eq:PowerSpectrumUx,eq:PowerSpectrumUy,eq:PowerSpectrumVort}) and the noise contributions (i.e. \cref{eq:NoiseUx,eq:NoiseUy,eq:NoiseVort}) yields our synthetic power spectrum model, in terms of azimuthal velocity, latitudinal velocity, and radial vorticity, respectively. Only two ingredients are needed to quantify these expressions, namely (i) the Green function $G(y,y_s)$ associated with the linear operator $\mathcal{L}$, and (ii) the turbulent stream function spectrum $\ep$.

\subsection{The Green function\label{subsec:GreensFunction}}

The angular frequency, $\omega$, and azimuthal wavenumber, $k_x$, being fixed, the linear operator, $\mathcal{L}$, depends on (i) the differential rotation profile, $U(y)$, (ii) the Coriolis parameter, $\beta = 2\oeq / R$, and (iii) the turbulent viscosity, $\nut$. Concerning the differential rotation profile, as a first step, we approximated it by a parabolic profile,
\begin{equation}
    U(\xi) = -\overline{U} \xi^2~, \qquad \xi \equiv y / R = \sin\lambda~,
    \label{eq:DifferentialRotation}
\end{equation}
where we have implicitly placed ourselves in a frame of reference rotating at the solar equatorial rotation rate $\oeq / (2\pi) = 453.1$ nHz, and we have introduced the non-dimensionalised latitudinal coordinate $\xi$. We chose the same value $\overline{U} = 244$ m~s$^{-1}$ as \citet{gizon20}, which approximates the solar differential rotation at low latitudes. With this value of $\oeq$, the Coriolis parameter becomes $\beta = 8.18 \times 10^{-15}$ m$^{-1}$s$^{-1}$. Finally, the turbulent viscosity, $\nut$, is specified through the turbulent Reynolds number,
\begin{equation}
    \ret \equiv \dfrac{\overline{U} R}{\nut}~,
    \label{eq:TurbulentReynoldsNumber}
\end{equation}
which we leave as a free parameter.

Once all these parameters are fixed, in order to numerically compute the Green function, we expand \cref{eq:DefinitionGreensFunction} on the basis formed by the Chebyshev polynomials of the first kind. Those are defined, for every positive integer $n$, by 
\begin{equation}
    T_n(\xi) = \cos(n\arccos \xi)~,
\end{equation}
which reduces to a polynomial expression after some algebraic manipulations. These polynomials are orthogonal to each other with respect to the following inner product,
\begin{equation}
    \bigbraket{f}{g} \equiv \displaystyle\int_{-1}^{1} \dfrac{f(\xi)g(\xi)}{\sqrt{1-\xi^2}} ~ \d \xi~,
\end{equation}
in the sense that
\begin{equation}
    \bigbraket{T_i}{T_j} = \dfrac{\pi}{2} c_i \delta_{ij}~,
\end{equation}
where $\delta_{ij}$ is the Kronecker delta and  $c_i = 1 + \delta_{i0}$. We denote the column vector containing the decomposition of the Green function on the Chebyshev basis as $\mathcal{G}(\xi_s)$, so that
\begin{equation}
    G(\xi,\xi_s) = \sum_{i=0}^\infty \mathcal{G}_i(\xi_s) T_i(\xi)~,
\end{equation}
where we also introduced the non-dimensionalised source position $\xi_s \equiv y_s / R$. The vector $\mathcal{G}(y_s)$ is the solution of the following linear system:
\begin{equation}
    \mathcal{M} \mathcal{G}(\xi_s) = \mathcal{B}(\xi_s)~,
    \label{eq:SymbolicLinearSystem}
\end{equation}
where the column vector on the right-hand side comprises the projections of the Dirac distribution on the Chebyshev polynomials,
\begin{equation}
    \mathcal{B}_i \equiv \dfrac{2}{\pi c_i} \bigbraket{T_i}{\delta(\xi - \xi_s)} = \dfrac{2T_i(\xi_s)}{\pi c_i \sqrt{1-\xi_s}}~,
    \label{eq:ColumnVectorGreen}
\end{equation}
and the matrix $\mathcal{M}$ on the left-hand side is defined by
\begin{equation}
    \mathcal{M}_{ij} \equiv \dfrac{2}{\pi c_i} \bigbraket{T_i}{\mathcal{L} T_j}~.
    \label{eq:LinearMatrixGreen}
\end{equation}
We note that the factor $2/(\pi c_i)$ in both Eqs. \ref{eq:ColumnVectorGreen} and \ref{eq:LinearMatrixGreen} stems from the fact that the set of Chebyshev polynomials is orthogonal but  not orthonormal.

The Chebyshev polynomials of the first kind prove particularly well suited for solving \cref{eq:DefinitionLinearOperator}, because the $\mathcal{M}_{ij}$ take a conveniently simple form on that basis, as was shown for example by \citet{orszag71}. We detail the derivation of these matrix coefficients in \cref{app:LinearMatrixGreen}. Naturally, while the matrix $\mathcal{M}$ is of infinite dimension, it is necessary to crop it to a finite size, in order for the numerical computations to be carried out. We found that truncating the Chebyshev expansion at $N = 500$ was a good compromise between a reasonable computation time and an accurate representation of the Green function. We note that while the right-hand side of \cref{eq:SymbolicLinearSystem} depends on the source position $\xi_s$, this is not the case of the matrix $\mathcal{M}$, meaning that for any given angular frequency $\omega$ and azimuthal wavenumber $k_x$, only one single $N \times N$ matrix inversion is necessary to find the Green function for all possible source positions.

Solving \cref{eq:SymbolicLinearSystem} for $\mathcal{G}(\xi_s)$ yields one Green function, corresponding to arbitrary and completely uncontrolled boundary conditions. It is therefore also necessary to enforce the correct boundary conditions:
\begin{align}
    & G\left(\xi=-1, \xi_s\right) = G\left(\xi=1, \xi_s\right) = 0~, \\
    & \left.\dfrac{\partial G}{\partial \xi} \right|_{\xi=-1,\xi_s} = \left.\dfrac{\partial G}{\partial \xi} \right|_{\xi=1,\xi_s} = 0~.
\end{align}
The Chebyshev polynomials of the first kind verify $T_i(1) = 1$, $T_i(-1) = (-1)^i$, $T'_i(1) = i^2$ and $T'_i(-1) = (-1)^{i+1} i^2$, so that enforcing these boundary conditions amounts to ensuring that the solution $\mathcal{G}(\xi_s)$ of \cref{eq:SymbolicLinearSystem} verifies
\begin{align}
    & \sum_{i=0}^{N-1} \mathcal{G}_i = \sum_{i=0}^{N-1} (-1)^i \mathcal{G}_i = 0~, \\
    & \sum_{i=0}^{N-1} i^2 \mathcal{G}_i = \sum_{i=0}^{N-1} (-1)^{i+1} i^2 \mathcal{G}_i = 0~.
\end{align}
These conditions can be enforced in \cref{eq:SymbolicLinearSystem} by replacing the last four lines of the matrix $\mathcal{M}$ by
\begin{align}
    & \mathcal{M}_{N-4,j} = 1~, \qquad \mathcal{M}_{N-3,j} = (-1)^j~, \\
    & \mathcal{M}_{N-2,j} = j^2~, \qquad \mathcal{M}_{N-1,j} = (-1)^{j+1} j^2~,
\end{align}
and by replacing the last four components of $\mathcal{B}(\xi_s)$ by zero. This is perfectly equivalent to the $\tau$-method applied, for example, by \citet{orszag71}. Stated more intuitively, this means that the high-frequency behaviour of the solution is now controlled not by the dynamical behaviour of the system, but by the mechanical constraints imposed on the boundaries. Solving this modified linear system for $\mathcal{G}(\xi_s)$ now yields the correct Green function, with the appropriate boundary conditions.

\subsection{The turbulent stream function spectrum\label{subsec:TurbulentSpectrum}}

The turbulent spectrum $\ep$, defined by \cref{eq:DefinitionStreamFunctionSpectrum}, is written in terms of the turbulent fluctuation of the stream function $\pt$. On the other hand, a more common definition for the turbulent spectrum relies on the turbulent velocity $\mathbf{u}_\mathrm{turb}$ rather than the stream function
\begin{equation}
    \eu(\omega, \mathbf{k}) \equiv \displaystyle\int \d\tau \d^2\mathbf{x} \left\langle u_{i,\mathrm{turb}}(T, \mathbf{X}) u_{j,\mathrm{turb}}(T+\tau, \mathbf{X}+\mathbf{x}) \right\rangle  \mathrm{e}^{\ii (\omega\tau - \mathbf{k} \cdot \mathbf{x})}~.
    \label{eq:DefinitionVelocitySpectrum}
\end{equation}
We note that, whether it be in \cref{eq:DefinitionStreamFunctionSpectrum} or in \cref{eq:DefinitionVelocitySpectrum}, the angular frequency $\omega$ is not restricted to be positive, but can be of any sign. These two spectra are related through
\begin{equation}
    \eu = \left(k^2 \delta_{ij} - k_i k_j \right) \ep~.
    \label{eq:RelationEuEp}
\end{equation}
The turbulent velocity spectrum is usually expressed as \citep[e.g.][]{lesieur08}
\begin{equation}
    \eu = \mathrm{E}(\mathbf{k}, \omega) \left(\delta_{ij} - \dfrac{k_i k_j}{k^2} \right)~.
    \label{eq:RelationEuEk}
\end{equation}
If the turbulence is incompressible, then the quantity $\mathrm{E}(\mathbf{k},\omega)$ is rigorously isotropic (in the sense that it does not depend on the direction of the wavevector $\mathbf{k}$, but only on its modulus), and the directional information is entirely contained in the projection operator that follows. This function $\mathrm{E}(\mathbf{k}, \omega)$ is what is commonly referred to as the turbulent spectrum. Plugging \cref{eq:RelationEuEk} into \cref{eq:RelationEuEp}, we simply find
\begin{equation}
    \ep(\omega, \mathbf{k}) = \dfrac{\mathrm{E}(\omega, \mathbf{k})}{k^2}~,
\end{equation}
so that knowing $\ep$ is perfectly equivalent to knowing $\mathrm{E}$.

\begin{figure*}
    \centering
    \begin{tabular}{cc}
    \includegraphics[width=0.5\linewidth]{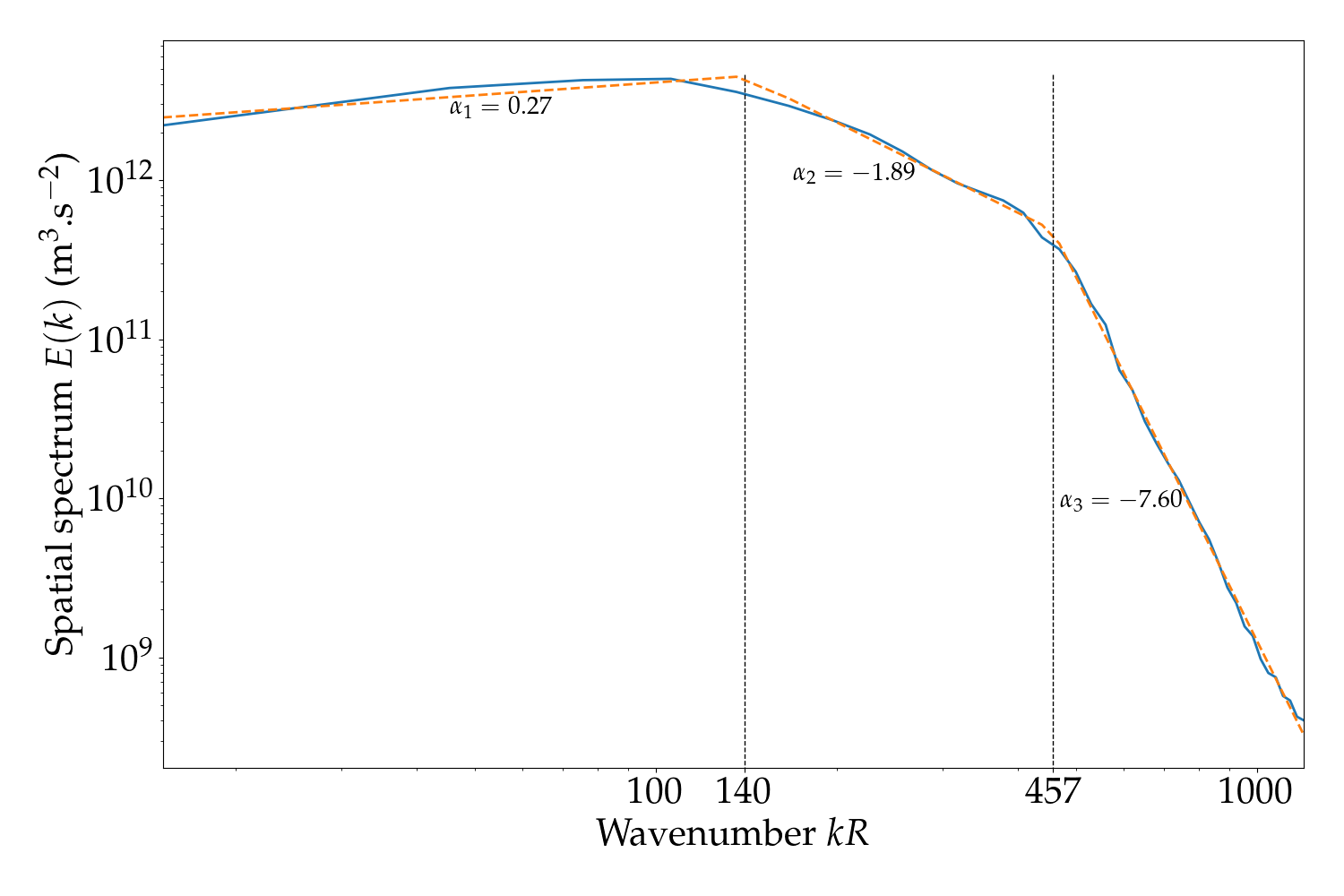} &
    \includegraphics[width=0.5\linewidth]{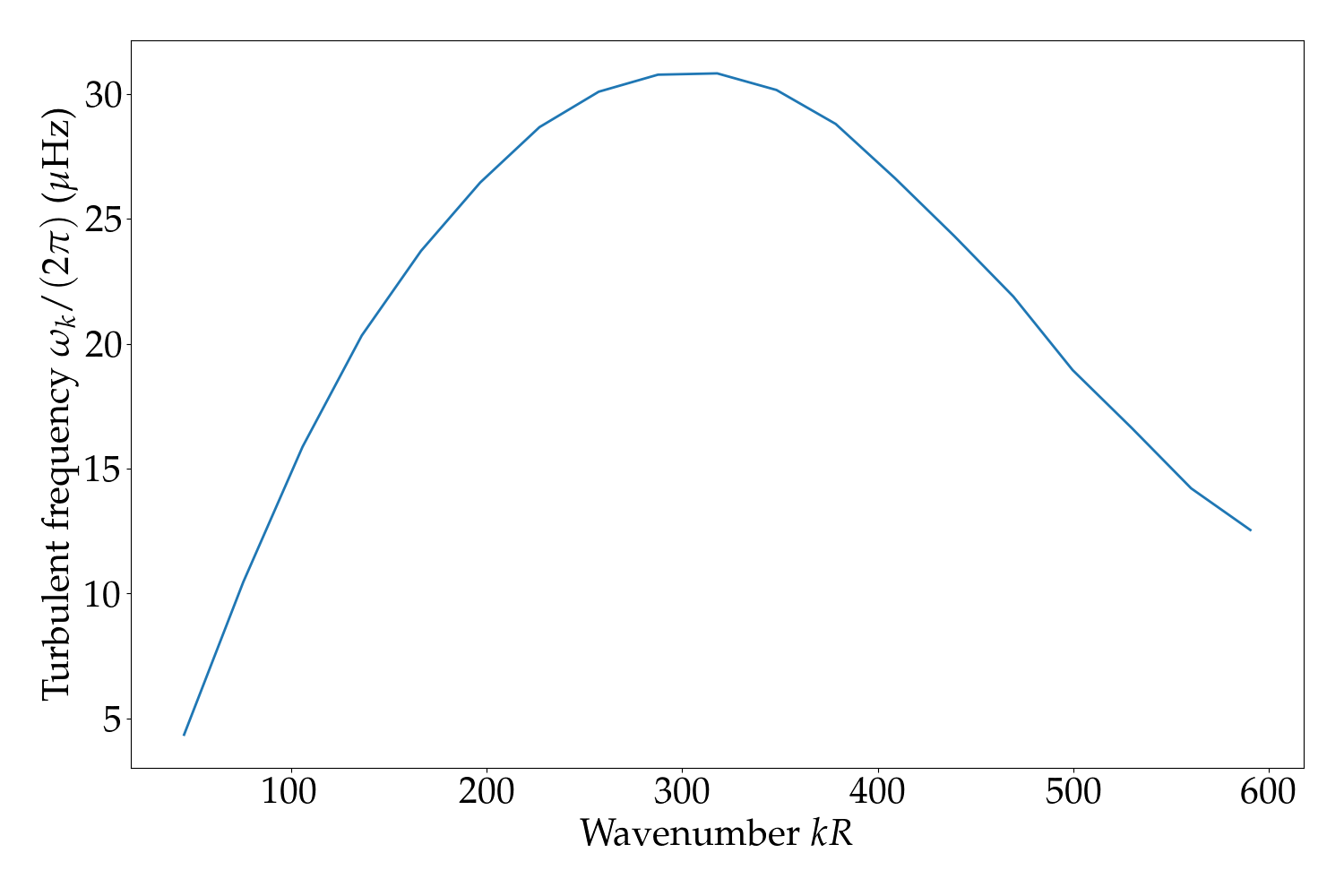}
    \end{tabular}
    \caption{Spatial turbulent spectrum and turbulent frequency. \textbf{Left:} Spatial turbulent spectrum, $\eiso(k),$ extracted from solar observations (solid blue line). The spectrum comprises roughly three sections, each with a different power law in $k^\alpha$. The dashed orange line shows the best fit of the data to a three-piece-wise power law. The exponents and the boundaries between the three regimes are indicated on the plot. \textbf{Right:} Turbulent frequency $\omega_k \equiv k u_k$ as a function of wavenumber, $k$, where $u_k$ is the typical velocity of the turbulent eddies of inverse size $k$, given by \cref{eq:VelocityScaleK}.}
    \label{fig:SpatialSpectrumTurbulentFrequencies}
\end{figure*}

In the following we consider that the turbulent spectrum $\mathrm{E}(\mathbf{k}, \omega)$ can be separated into a spatial part and a temporal part. The argument was proposed by \citet{stein67} and later consistently used in the context of wave excitation by turbulent emission (see for instance \citealt{musielak94}, or, in the context of solar-like $p$-modes, \citealt{samadi01,chaplin05}; see also \citealt{tennekes78} Chapter 8 for a textbook on the subject). The idea is to introduce the spatial spectrum
\begin{equation}
    E(\mathbf{k}) \equiv \displaystyle\int_{-\infty}^{+\infty} \d\omega ~ \mathrm{E}(\mathbf{k},\omega)~
\end{equation}
and then define the temporal part of the spectrum as

\begin{equation}
    \chi_{\mathbf{k}}(\omega) \equiv \dfrac{\mathrm{E}(\mathbf{k},\omega)}{E(\mathbf{k})}~.
    \label{eq:DefinitionChik}
\end{equation}
We consider that, as in the case of incompressible turbulence, the spatial spectrum $E(\mathbf{k})$ is isotropic, and can therefore be rewritten as a kinetic energy per unit $k$ instead of per unit $\mathbf{k}$,
\begin{equation}
    E(\mathbf{k}) = \dfrac{\eiso(k)}{2\pi k}~.
\end{equation}

Concerning the temporal spectrum, following dimensional arguments, \citet{stein67} argued that the temporal evolution of a turbulent eddy of size $\lambda_k \equiv 2\pi / k$ should contain frequencies around $\nu_k \sim u_k / \lambda_k$ (or, equivalently, angular frequencies around $\omega_k \sim k u_k$), where $u_k$ is the typical velocity associated with these eddies. Guided by the work of \citet{kraichnan65} (see his equation 9.6 and the discussion in the paragraph above it), \citet{stein67} wrote
\begin{equation}
    \chi_{\mathbf{k}}(\omega) = \dfrac{1}{\omega_k} \chi(\widetilde{\omega})~, \qquad \widetilde{\omega} \equiv \omega / \omega_k~,
    \label{eq:RescalingTemporalSpectrum}
\end{equation}
where we have introduced the reduced frequency $\widetilde{\omega}$, and $\chi$ is now a function of frequency that does not depend on the eddy size $k$. \citet{stein67} proved that the typical velocity $u_k$ associated with these eddies is determined by the spatial spectrum through
\begin{equation}
    u_k \equiv \left( \displaystyle\int_{k < k' < 2k} \d k' ~ \eiso(k') \right)^{1/2}~.
    \label{eq:VelocityScaleK}
\end{equation}

All in all, we can write
\begin{equation}
    \ep = \dfrac{\eiso(k) ~ u_k ~ \chi(\widetilde{\omega})}{2\pi k^2}~,
\end{equation}
where $u_k$ is given by \cref{eq:VelocityScaleK}. Describing the whole turbulent spectrum requires only two ingredients, namely (i) the $\omega$-independent spatial spectrum $\eiso$ and (ii) the $k$-independent temporal spectrum $\chi$. Both can be extracted from solar observations. We used measurements of the vertical vorticity deduced from granulation tracking by \citet{langfellner15} (their Fig. 3, where the units where reassessed). We use the observations near the solar equator and adopt the same vorticity spectrum at all latitudes, for the sake of simplicity.

The spatial spectrum $\eiso(k)$ is shown in the left panel of \cref{fig:SpatialSpectrumTurbulentFrequencies}, and can clearly be described by three distinct power laws in three separate wavenumber regimes:
\begin{equation}
    \eiso(k) =
    \begin{cases}
        C_1 \left(\dfrac{k}{k_\mathrm{ref}}\right)^{\alpha_1} & \mathrm{if} ~~ k \le k_1~, \\
        C_2 \left(\dfrac{k}{k_\mathrm{ref}}\right)^{\alpha_2} & \mathrm{if} ~~ k_1 < k < k_2~, \\
        C_3 \left(\dfrac{k}{k_\mathrm{ref}}\right)^{\alpha_3} & \mathrm{if} ~~ k_2 \le k~,
    \end{cases}
\end{equation}
where the choice of $k_\mathrm{ref}$ is completely arbitrary and can be absorbed in the factors $C_i$. The spectrum is therefore parameterised by (i) the three exponents $\alpha_{1,2,3}$, (ii) the two wavenumber cutoffs $k_1$ and $k_2$, and (iii) the total turbulent kinetic energy per unit mass $\int \eiso(k) \d k$. The best fit to the observational data is also shown in the left panel of \cref{fig:SpatialSpectrumTurbulentFrequencies}. We find exponents $\alpha_1 = 0.27$, $\alpha_2 = -1.88$ and $\alpha_3 = -7.60$; in particular, the slope $\alpha_2$ of the middle section is quite close to the $-5/3$ power law theoretically predicted for the inertial range under the Kolmogorov hypotheses. As for the wavenumber cutoffs -- also indicated in the left panel of \cref{fig:SpatialSpectrumTurbulentFrequencies} -- we find $k_1 R = 140$ and $k_2 R = 457$. These values are significantly larger than the wavenumber associated with the solar inertial modes, thus validating the assumption made earlier concerning the scale separation.

We then use this spatial spectrum to compute the turbulent frequencies $\omega_k = k u_k$ as a function of $k$, where $u_k$ is given by \cref{eq:VelocityScaleK}. Those are shown in the right panel of \cref{fig:SpatialSpectrumTurbulentFrequencies}. It can be seen that the typical frequencies associated with the turbulent motions are of the order of a few tens of $\mu$Hz. The frequencies of the inertial modes, by contrast, are typically much lower (of the order of $\sim 100$ nHz), which means that there is a timescale separation between the inertial modes and the turbulence, similar to the length scale separation already mentioned above.

\begin{figure}
    \centering
    \includegraphics[width=\linewidth]{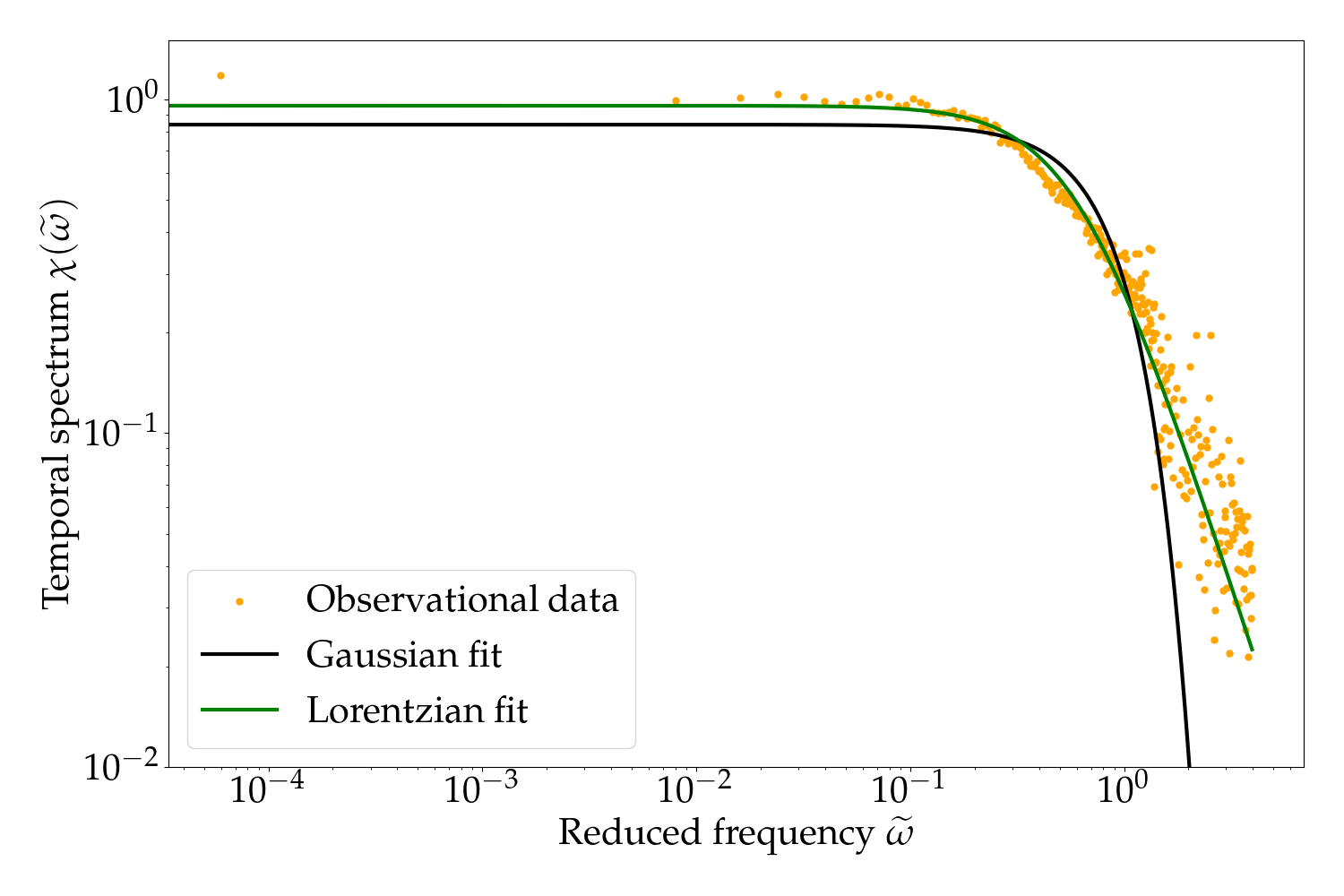}
    \caption{Temporal spectrum $\chi(\widetilde{\omega}) \equiv \omega_k ~ \chi_k(\omega)$, as a function of the reduced frequency $\widetilde{\omega} \equiv \omega / \omega_k$. The data points (in orange) have been binned according to the value of $\widetilde{\omega}$ (which depends on both $\omega$ and $k$), and only the mean over each bin is shown. These data points collapse onto a unique, slowly varying curve, which was adjusted alternatively with a Gaussian function (\cref{eq:GaussianFit}, solid black line) and a Lorentzian function (\cref{eq:LorentzianFit}, solid green line). The latter is clearly the best fit, obtained for an amplitude $A = 1.85$ and a standard deviation $\sigma = 0.62$.}
    \label{fig:TemporalSpectrum}
\end{figure}

Finally, we checked the assumption made in writing \cref{eq:RescalingTemporalSpectrum} (i.e. the fact that the quantity $\omega_k \chi_k$, when plotted against the reduced frequency $\widetilde{\omega}$, collapses onto a unique, slowly varying curve, independent of $k$). The result, shown in \cref{fig:TemporalSpectrum}, seems to indicate that this is indeed the case. What is more, it is also possible to determine the analytical function that best describes this curve. In the context of $p$-mode excitation, traditional models for the turbulent temporal spectrum usually assume either a Gaussian or a Lorentzian function \citep[e.g.][]{goldreich77b,balmforth92c,samadi07,belkacem10}. We tried the two following models:
\begin{equation}
    \chi(\widetilde{\omega}) = \dfrac{A}{\sqrt{2\pi \sigma^2}} e^{-\widetilde{\omega}^2 / (2 \sigma^2)}
    \label{eq:GaussianFit}
\end{equation}
and
\begin{equation}
    \chi(\widetilde{\omega}) = \dfrac{A}{\pi\sigma} \dfrac{1}{1 + \left( \widetilde{\omega} / \sigma \right)^2}~.
    \label{eq:LorentzianFit}
\end{equation}
In each case, the dimensionless factor $\sigma$ is introduced to account for the uncertainty on the relation $\omega_k = k u_k$. It constitutes a free parameter in each fit, but is expected to be of order unity. This parameter is akin to the parameter $\lambda$ in the $p$-mode excitation formalism of \citet{samadi01}, which they also left free for the same reason. \Cref{fig:TemporalSpectrum} shows the result of each fitting procedure. The Lorentzian function clearly yields the best agreement, with an amplitude $A = 1.85$ and standard deviation $\sigma = 0.62$; we therefore adopted this prescription. We note that this value of $\sigma$ is consistent with the values of $\lambda$ found by \citet{samadi01}, who constrained it using the observed excitation rate of solar acoustic modes (see their Table 1).

%%% Section 3

\section{Results\label{sec:Results}}

\subsection{Equatorial power spectrum\label{subsec:EquatorialSpectra}}

We first considered the synthetic spectrum as it would be observed at the solar equator, and written in terms of latitudinal velocity $u_y$. This is obtained by setting $\yo = 0$ in \cref{eq:PowerSpectrumUy} and \cref{eq:NoiseUy}, which yield respectively the contribution of the inertial modes and of the turbulent noise to the overall spectrum. For the moment, however, we only consider the inertial mode contribution, and we only add the turbulent noise contribution later on. The power spectrum thus obtained corresponds to a spectral density per unit longitudinal wavevector $k_x$ and angular frequency $\omega$. It is then straightforward to transform it into a power spectral density per unit $\omega$ only, for any given azimuthal order $m \equiv k_x R$, by dividing the power spectrum by the radius, $R$, of the spherical domain. Naturally, our model can be applied to any real value of $m$, because in the scope of the equatorial $\beta$-plane approximation we did not make any assumption regarding the $\phi$-periodicity of the velocity field. However, for the sake of comparison with observations, only integer values of $m$ are relevant.

\subsubsection{Individual mode contributions\label{subsubsec:ModeBreakdown}}

\begin{figure}[h!]
    \centering
    \includegraphics[width=\linewidth]{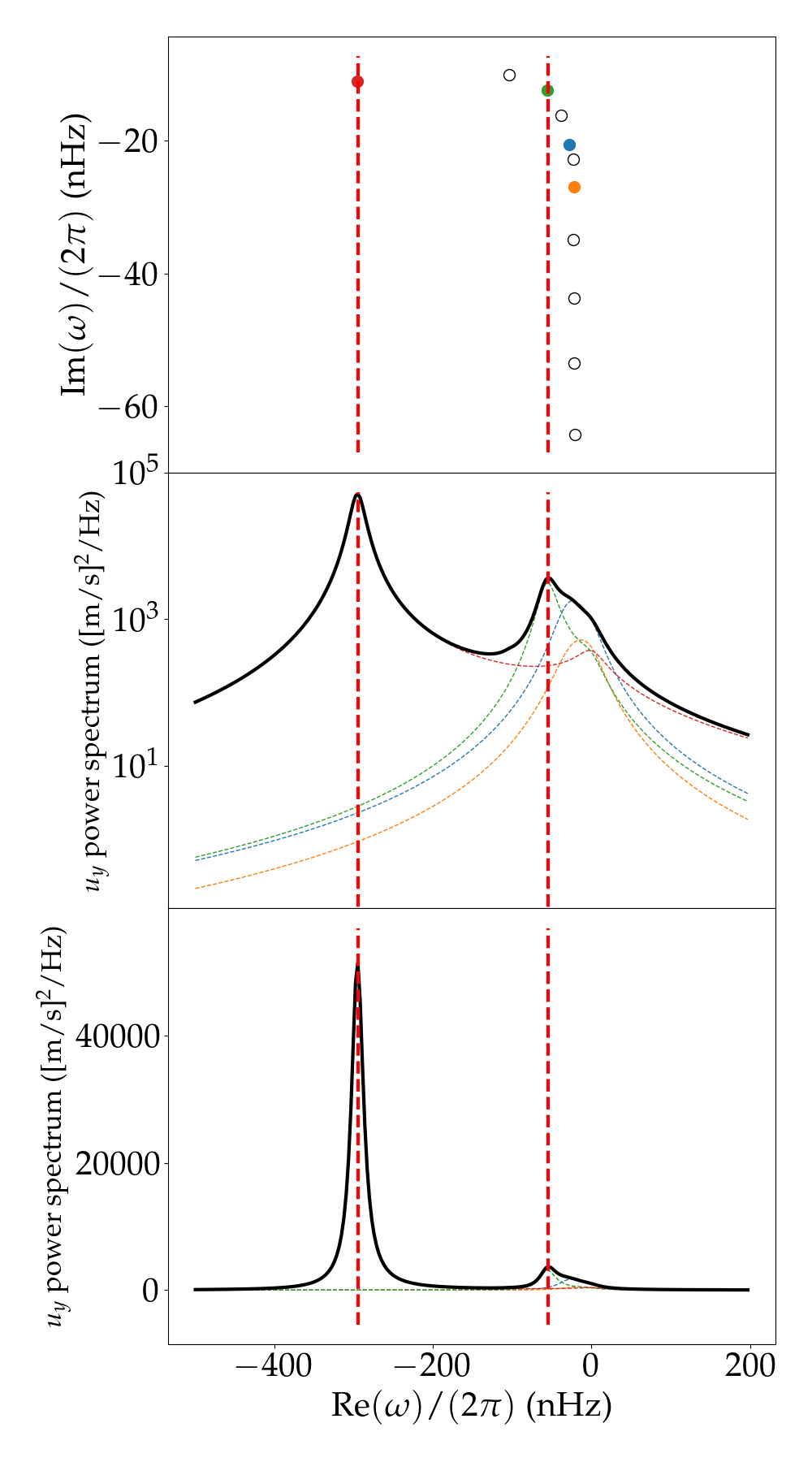}
    \caption{Latitudinal velocity spectrum. \textbf{Top:} Discrete inertial mode eigenspectrum for $m = 3$, shown in the complex plane. Each point correspond to one mode: all eigenfrequencies have a negative real part (meaning the modes are retrograde) and a negative imaginary part (meaning they are stable). The coloured dots mark the modes whose contribution to the $u_y$ spectrum is most prominent. The vertical dashed lines indicate the central frequencies of each resonant peak in the $u_y$ power spectrum (see bottom panel). \textbf{Middle:} Equatorial $u_y$ spectrum, obtained from \cref{eq:PowerSpectrumUy} for azimuthal order $m = 3$. The solid black line shows the total spectrum, while each coloured dashed line corresponds to the individual contribution of the normal eigenmodes of the system, as described in the main text. The colours of the dashed lines match the colour scheme of the top panel. The red vertical dashed lines show the local maxima of the total spectrum. \textbf{Bottom:} Same as the middle panel, but the vertical scale is linear.}
    \label{fig:EqSpecM03}
\end{figure}

\begin{figure}
    \centering
    \includegraphics[width=\linewidth]{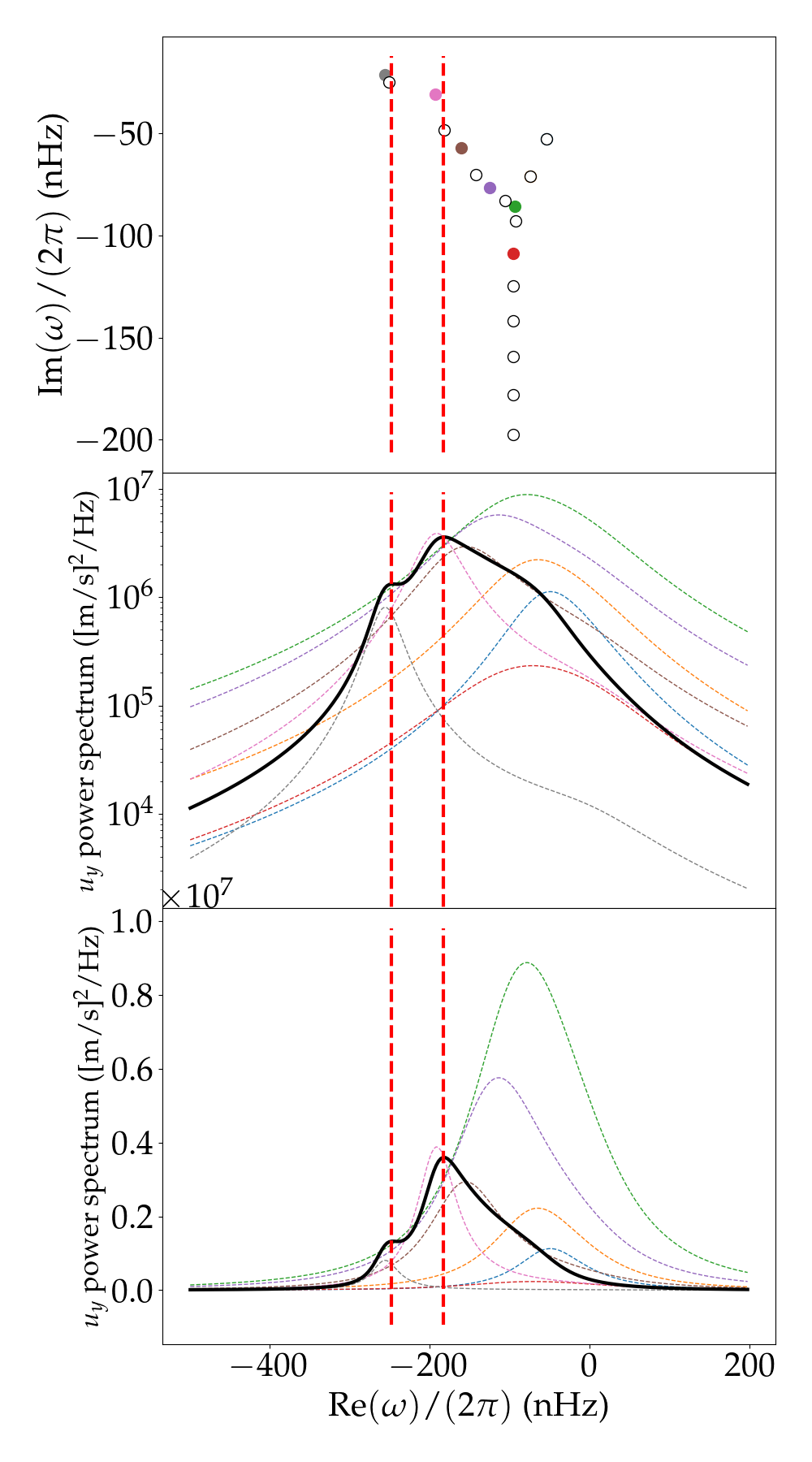}
    \caption{Same as \cref{fig:EqSpecM03}, but for $m = 5$.}
    \label{fig:EqSpecM05}
\end{figure}

\begin{figure}
    \centering
    \includegraphics[width=\linewidth]{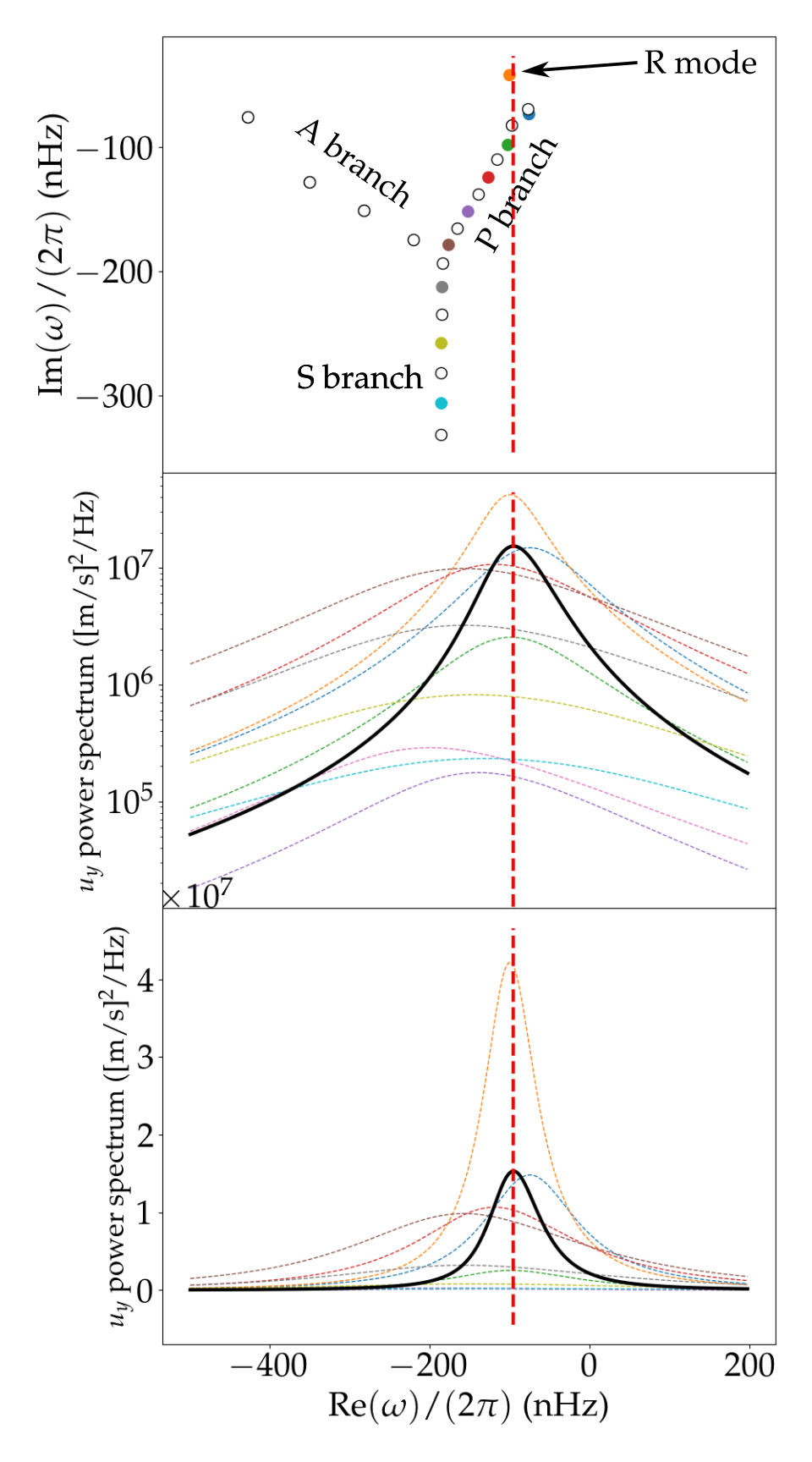}
    \caption{Same as \cref{fig:EqSpecM03}, but for $m = 10$. The three classical branches of Poiseuille flows are indicated on the plot, in addition to the R mode.}
    \label{fig:EqSpecM10}
\end{figure}

Our model allows us to decompose the total spectrum into the individual contributions of each inertial mode. To that effect, we first needed to compute the complex eigenfrequencies and eigenfunctions of the linear homogeneous system $\mathcal{L}\Psi = 0$, where we recall that $\mathcal{L}$ is given by \cref{eq:DefinitionLinearOperator}. The eigenfrequencies $\omega_n$ and eigenfunctions $\Psi_n$ are the solution of the following generalised boundary eigenvalue problem
\begin{equation}
    \mathcal{L}_1 \Psi_n = \omega_n \widehat{\Delta}\Psi_n~,
    \label{eq:GeneralisedEigenvalueProblem}
\end{equation}
where
\begin{equation}
    \mathcal{L}_1 \equiv k_x U \widehat{\Delta} + k_x(\beta - U'') + \ii\nut\widehat{\Delta}^2~.
\end{equation}
The enforced boundary conditions are the same as those of the inhomogeneous problem (see Sect. \ref{subsec:GreensFunction}). We solve the system numerically, as described in \cref{subsec:GreensFunction}, by projecting the generalised boundary eigenvalue problem on the basis formed by the Chebyshev polynomials of the first kind. The discrete eigenspectrum is showcased in the complex plane, for azimuthal orders $m = 1$, $5$ and $10$ in the top panels of \cref{fig:EqSpecM03,fig:EqSpecM05,fig:EqSpecM10}, where each point represents one mode. As previously mentioned, all the eigenfrequencies have a negative real part, meaning that the modes propagate in the retrograde direction, and a negative imaginary part, meaning that they are all stable, and none of them is exponentially growing. In the $m = 5$ and $m = 10$ cases, one can clearly recognise the three classical mode branches associated with Poiseuille flows \citep{mack76}. Those are indicated in the top panel of \cref{fig:EqSpecM10}. While the two upper branches (i.e. the A branch and the P branch) comprise a finite number of modes, all of which are shown, the vertical branch (i.e. the S branch) is infinite, and is truncated in the plots. Furthermore, in those same cases, an additional mode sticks out (it is marked as R mode in the top panel of \cref{fig:EqSpecM10}), characterised by a comparatively longer lifetime (i.e. an imaginary frequency closer to zero). It is entailed by the presence of global rotation in the system (through the $\beta$ factor in \cref{eq:DefinitionLinearOperator}), and can be thought of as the equivalent of an equatorial Rossby mode in Cartesian geometry. Overall, the structure of the eigenspectrum is the same as the one presented in \citet{gizon20}, as the homogeneous part of our wave equation is the same as theirs; for more details on the matter, we therefore refer the reader to their work.

We also need to compute the complex eigenfrequencies and eigenfunctions of the associated adjoint system $\mathcal{L}^\dagger\Psi = 0$, where the adjoint linear operator $\mathcal{L}^\dagger$ is defined in such a way that for any function $f$ and $g$ satisfying the boundary conditions of the problem, we have
\begin{equation}
    \displaystyle\int_{-1}^1 \d \xi ~ \left(\mathcal{L}^\dagger g\right)^\ast f = \displaystyle\int_{-1}^1 \d \xi ~ g^\ast \left(\mathcal{L} f\right)~.
\end{equation}
Performing integration by parts we find that the adjoint eigenfrequencies $\omega_n^\dagger$ and eigenfunctions $\Psi_n^\dagger$ are the solution of the following generalised boundary eigenvalue problem \citep{salwen81}
\begin{equation}
    \mathcal{L}_1^\dagger \Psi_n^\dagger = \omega_n^\dagger \widehat{\Delta} \Psi_n^\dagger~,
    \label{eq:GeneralisedAdjointEigenvalueProblem}
\end{equation}
where
\begin{equation}
    \mathcal{L}_1^\dagger \equiv k_x U \widehat{\Delta} + k_x \left(\beta + 2 U' \dfrac{\d}{\d \xi} \right) - \ii\nut\widehat{\Delta}^2~.
\end{equation}
It is easy to show that $\omega_n^\dagger = \omega_n^\ast$. Furthermore,  the eigenfunctions and adjoint eigenfunctions form a biorthogonal set, in the sense that
\begin{equation}
    \displaystyle\int_{-1}^1 \d \xi ~ \left( \dfrac{\d \Psi_n^\ast}{\d y} \dfrac{\d \Psi_m^\dagger}{\d y} + k_x^2 \Psi_n^\ast \Psi_m^\dagger \right) \propto \delta_{nm}~.
    \label{eq:Biorthogonality}
\end{equation}
The eigenfunctions are normalised in such a way that the proportionality coefficient in \cref{eq:Biorthogonality} is unity. This biorthonormality relation allows us to project the Green function $G(\xi,\xi_s)$ onto each of the modes alternatively, and therefore to compute the $u_y$ spectrum associated with each mode separately.

The results are shown in the bottom panels of \cref{fig:EqSpecM03,fig:EqSpecM05,fig:EqSpecM10}, for azimuthal orders $m = 3$, $5,$ and $10$, respectively. The total spectrum is represented by the solid black line, while the coloured dashed lines represent the individual contributions of the most prominent modes in the discrete spectrum. We note that for the most part, and as is expected of a resonant mode whose driving source has a broadband frequency dependence, the spectrum contribution of each individual mode takes the form of a Lorentzian profile. For low values of $m$ (more specifically for $m \leqslant 5$), the spectrum is clearly dominated by two main peaks, and the frequencies of these peaks makes them clearly identifiable, as they fall very close to the real part of discrete eigenmodes of the homogeneous system, as shown in the top panels.

As $m$ increases, the peaks become wider (because the imaginary part of the eigenfrequencies increase in modulus), and for $m \geqslant 6$ the spectrum becomes dominated by one mode only. From the eigenspectrum shown in the top panels it can be seen that the dominant mode corresponds to the equatorial Rossby mode. Interestingly, we find that several modes have an amplitude that, if taken individually, should lead to a visible peak in the spectrum, but remain invisible in the total spectrum, where all modes are accounted for at once. This seems to suggest that the reason these modes cannot be extracted from the spectrum is not the inefficiency of the excitation process but rather the fact that they form a mutually destructive interference pattern with each other. This is possible because the modes all share the same driving source. This interference phenomenon can only occur between modes that share the same azimuthal order $m$, because the problem is axisymmetric, and therefore different $m$ are completely decoupled. Furthermore, the modes must share similar eigenfrequencies (more specifically, the frequency difference must be of the order or smaller than the inverse of their lifetime).

\subsubsection{Frequencies, amplitudes, and linewidths\label{subsubsec:AmplitudesLinewidths}}

\begin{figure}[ht!]
    \centering
    \begin{tabular}{c}
        \includegraphics[width=\linewidth]{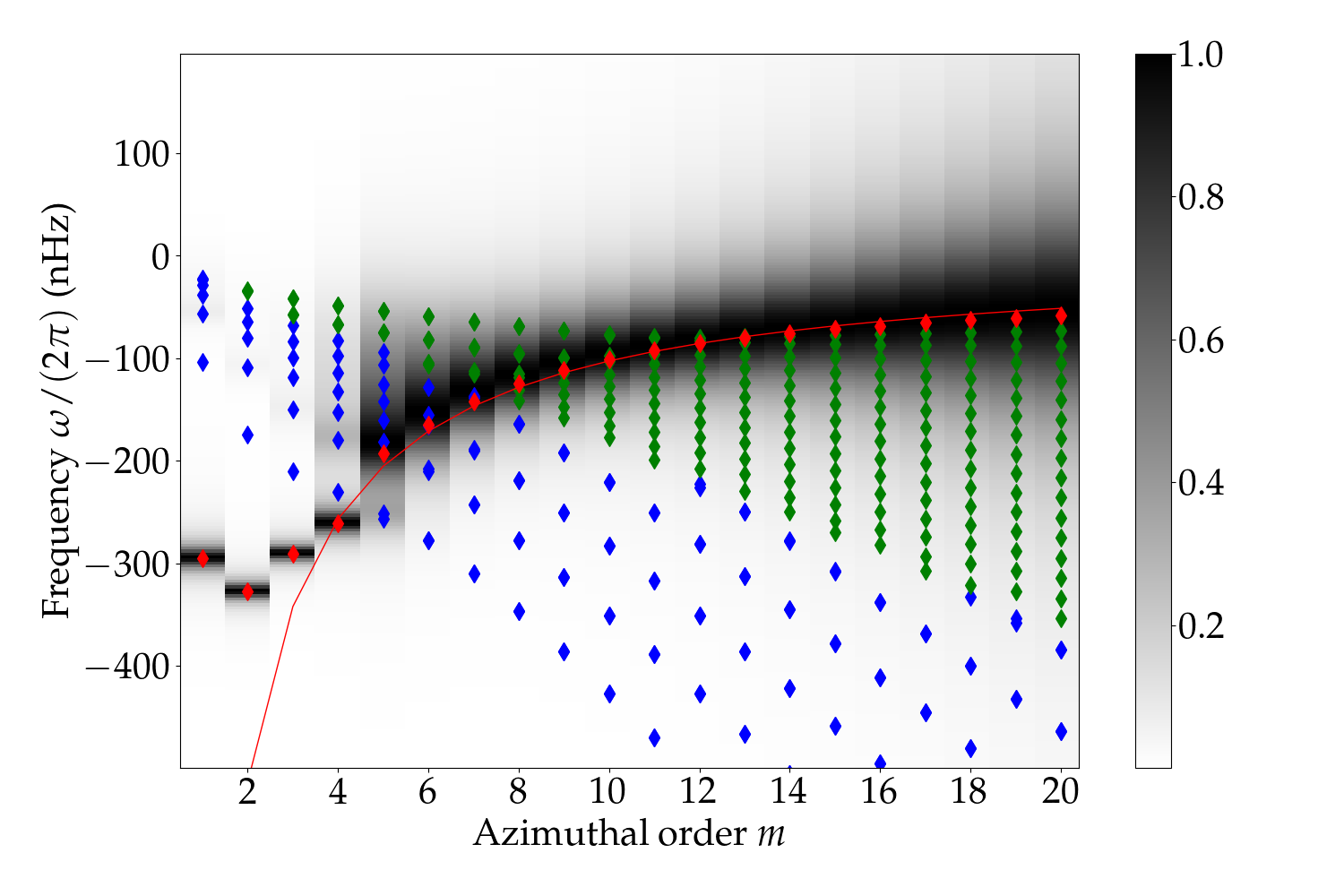} \\
        \includegraphics[width=\linewidth]{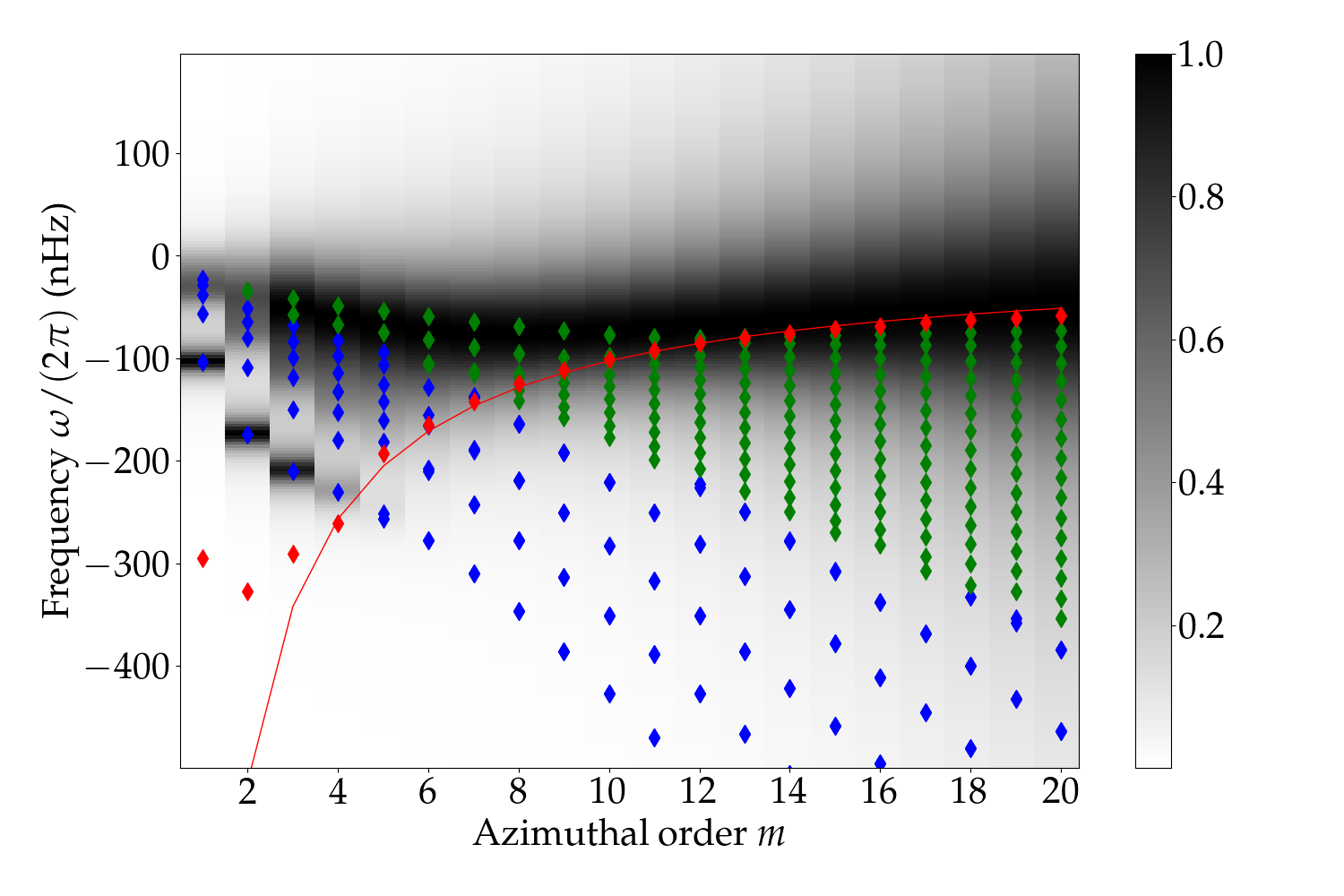} \\
        \includegraphics[width=\linewidth]{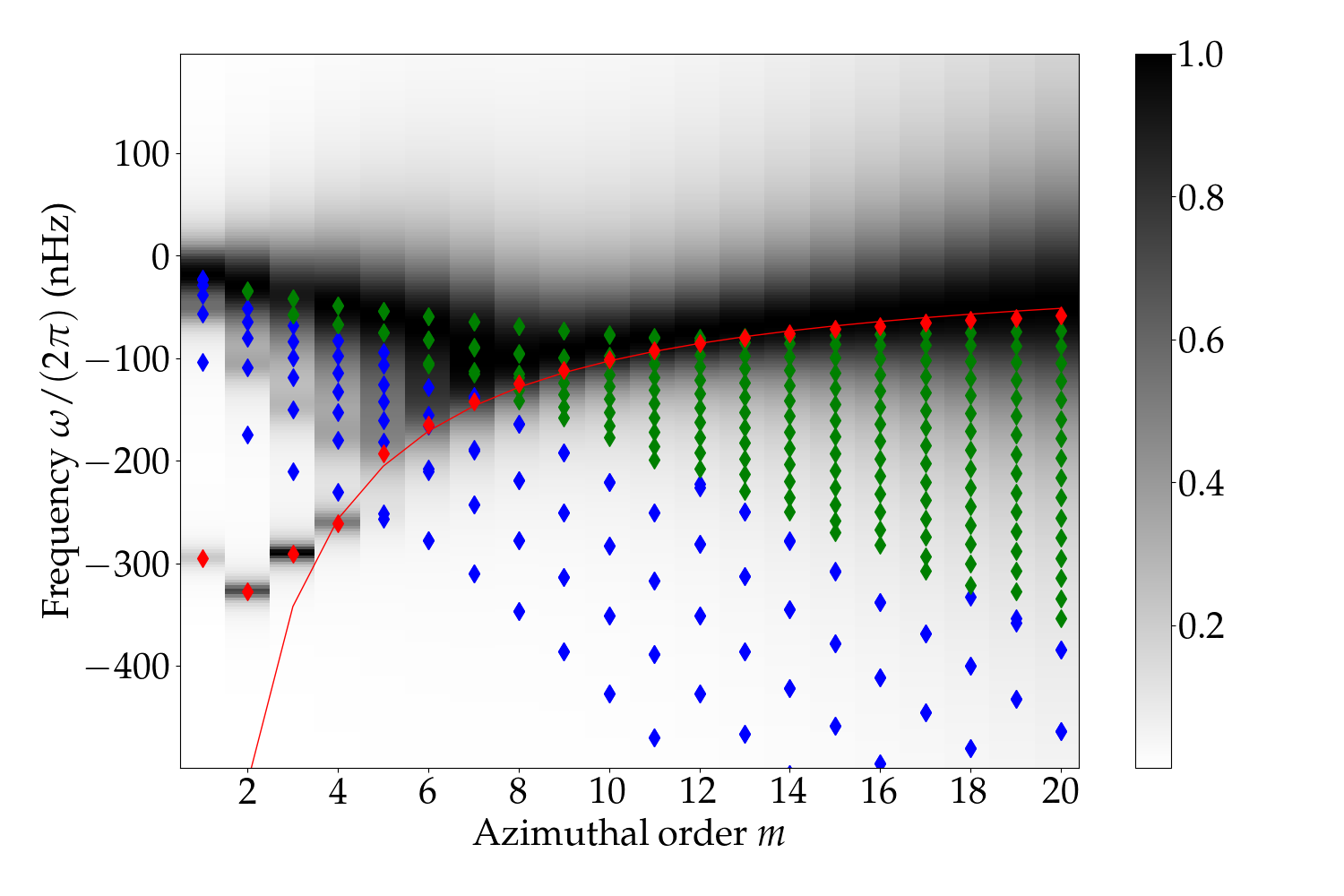}
    \end{tabular}
    \caption{Synthetic equatorial spectrum in the $m$-$\omega$ plane, in terms of $u_y$ (\textbf{top}), $u_x$ (\textbf{middle}), and $\zeta$ (\textbf{bottom}). Each vertical slice is normalised separately such that the maximum is unity. The diamonds show the real part of the eigenfrequencies of the linear homogeneous problem, computed as described in Sect. \ref{subsubsec:ModeBreakdown}. The colour code refers to the mode categories: the blue diamonds represent the A branch, the green diamonds represent the P branch, and the red diamonds represent the Rossby modes (see Sect. \ref{subsubsec:ModeBreakdown} for a description of these branches). The solid red line shows the theoretical dispersion relation for sectoral Rossby modes in Cartesian coordinates (see Eq. \ref{eq:RossbyDispersionRelation}).}
    \label{fig:Frequencies}
\end{figure}

For the resonant peaks that are sufficiently separated from each other in the synthetic spectrum, it is possible to directly infer, from the model, not only their frequency, but also their amplitude and linewidths. We define the angular frequency $\omega_0$ of each peak as the location of their local maximum, and their full linewidth at half maximum $\Gamma$ as the angular frequency range where the power spectral density is above half the height of the peak. The amplitude is defined as
\begin{equation}
    A = \left( \displaystyle\int_{\omega_a}^{\omega_b} \d\omega ~ P(\omega) \right)^{1/2}~,
    \label{eq:DefinitionAmplitude}
\end{equation}
where the boundaries $\omega_a$ and $\omega_b$ should be chosen to enclose most of the peak, without overlap with the other peaks. In practice, we chose $\omega_{a,b} = \omega_0 \mp \Gamma / 2$. In the case of a Lorentzian profile, this range encloses exactly half the energy of the mode, so we only have to apply a factor of$\sqrt{2}$ to Eq. \ref{eq:DefinitionAmplitude} to find the total mode amplitude. We note that, because of the typical linewidths of the modes, the upper boundary $\omega_b$ can very well be positive, despite the fact that the central frequency, $\omega_0$, is systematically negative.

The frequencies of the modes are showcased, as a function of $m$, in Fig. \ref{fig:Frequencies}. The coloured diamonds are identical on each panel: blue and green symbols represent each of the upper mode branches in the spectrum (see Sect. \ref{subsubsec:ModeBreakdown} for a description of these categories), while red symbols represent equatorial Rossby modes. On the background of each panel is superimposed an image of the spectrum in terms of different physical quantities (latitudinal velocity on top, azimuthal velocity in the middle, and vorticity at the bottom), in the $m$-$\omega$ plane, where each vertical slice is normalised so that the maximum is unity. One can distinguish the same transition, already mentioned above, between low azimuthal orders, for which several clearly identifiable resonant peaks can be resolved in the spectrum, and high azimuthal orders, where the distinction is no longer as clear. The theoretical Rossby mode dispersion relation is also plotted in each panel: in Cartesian coordinates, and in the presence of a background azimuthal jet $U$, it is given by
\begin{equation}
    \omega_R = -\dfrac{\beta - U''}{k_x}~.
    \label{eq:RossbyDispersionRelation}
\end{equation}
The high m frequencies match the theoretical dispersion relation quite well. The agreement, however, is not as close for lower values of m. We also note that while the low-m equatorial Rossby modes are quite distinguishable in the $u_y$ spectrum or the vorticity equatorial spectrum, they do not show in the equatorial $u_x$ spectrum, due to the fact that their $u_x$ eigenfunction has a node at the equator.

\begin{figure*}
    \centering
    \begin{tabular}{cc}
        \includegraphics[width=0.5\linewidth]{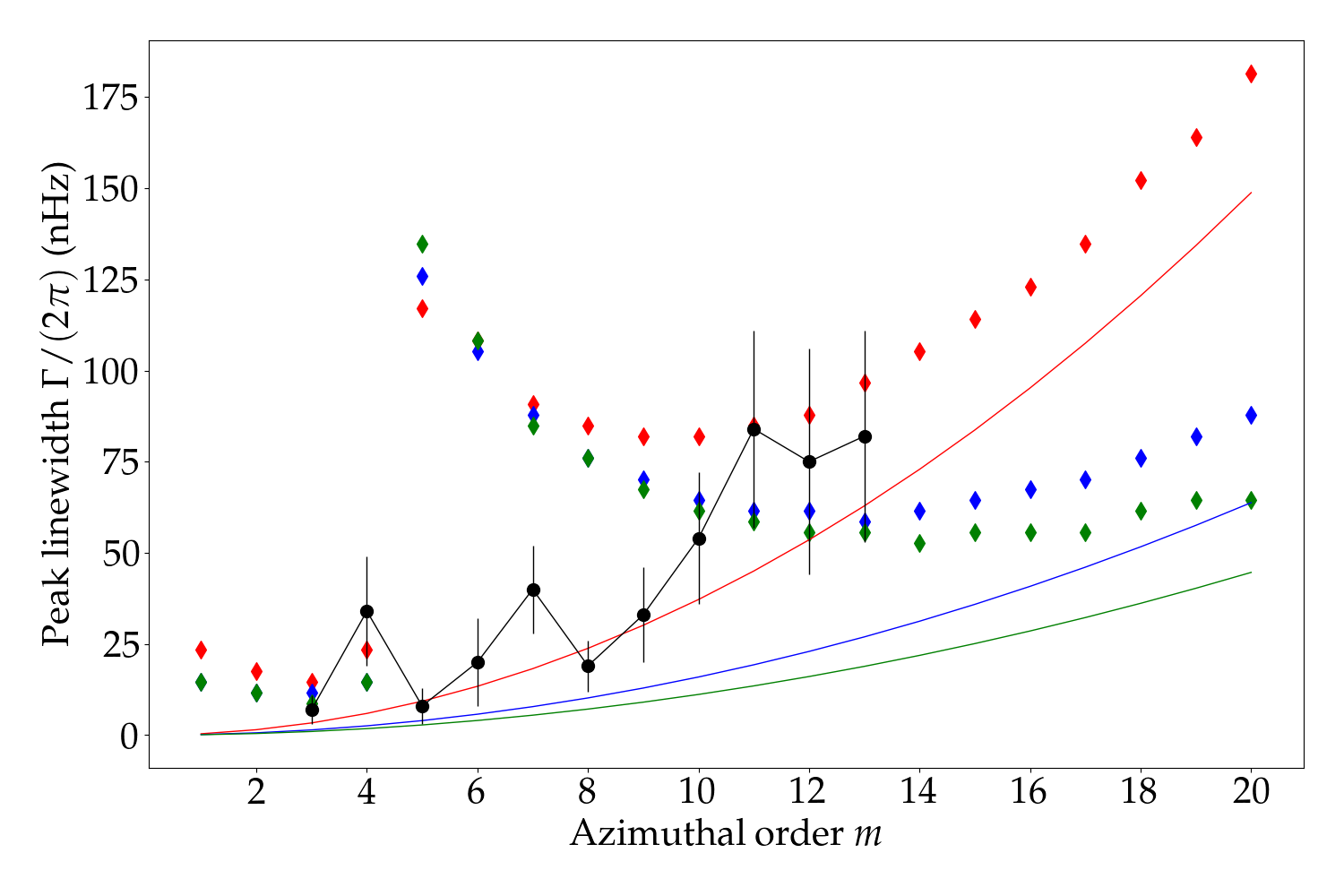} &
        \includegraphics[width=0.5\linewidth]{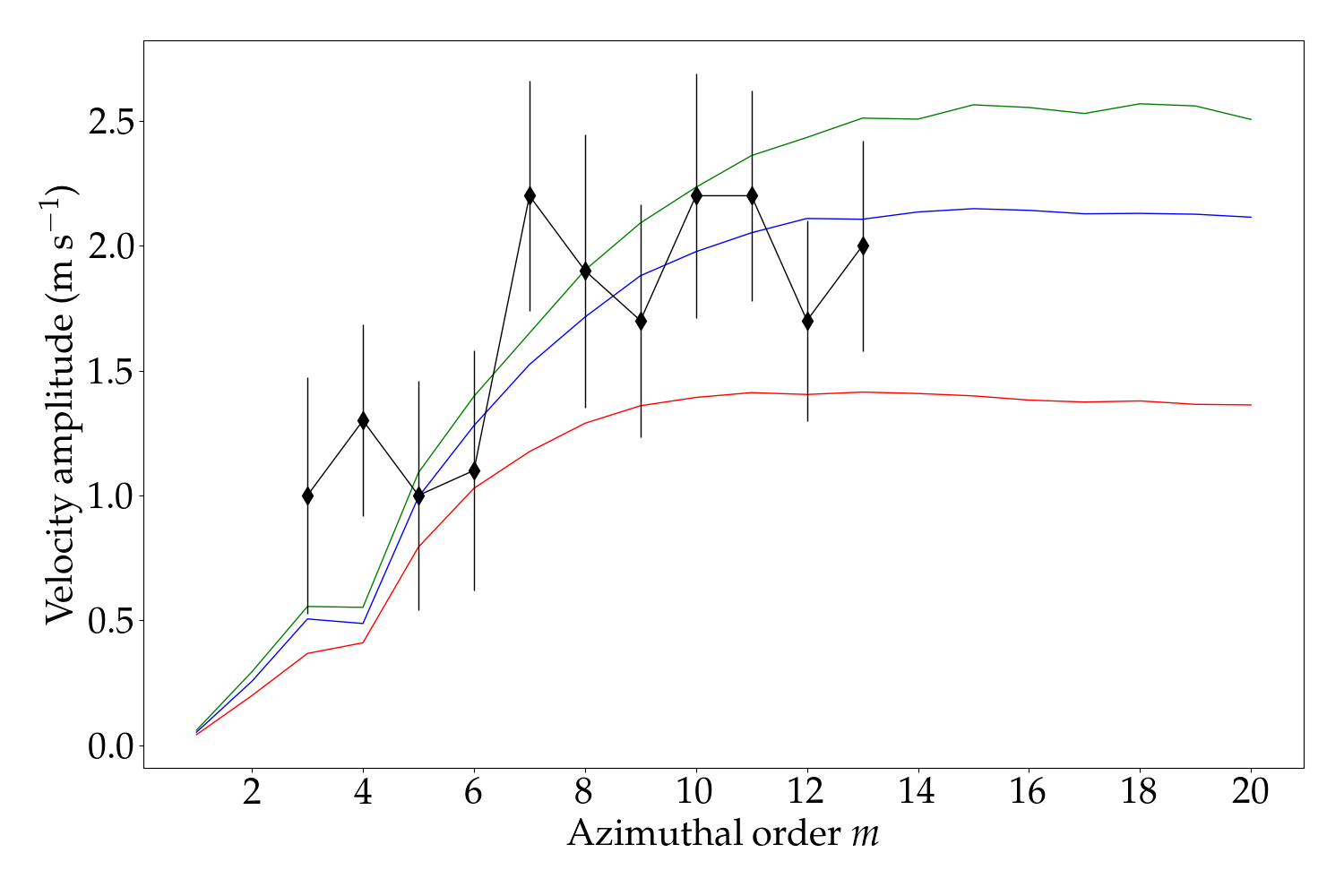}
    \end{tabular}
    \caption{Equatorial Rossby mode parameters. \textbf{Left:} Full width at half maximum of the resonant peaks that could be identified as equatorial Rossby modes, as a function of $m$. The diamonds show the linewidths measured in the $u_y$ equatorial synthetic power spectrum, and the coloured solid lines represent the theoretical Rossby mode linewidth, obtained from the classical dispersion relation $\Gamma_R = \nut k_x^2$ (see Eq. \ref{eq:RossbyTheoreticalLinewidth}). The colour code refers to the value of the turbulent Reynolds number: $\ret = 300$ (red), $700$ (blue), and $1000$ (green). The black line shows the mode linewidths inferred from solar observations at the equator in the latitudinal velocity spectrum, as reported by \citet{liang19}. Error bars from the fitting procedure reported by the authors are also shown. \textbf{Right:} Rossby mode amplitude (coloured solid lines) in the $u_y$ equatorial synthetic power spectra, as a function of azimuthal order $m$, defined as described in the text (see Eq. \ref{eq:DefinitionAmplitude}). The colour code is identical to the one in the left panel.}
    \label{fig:LinewidthsAmplitudes}
\end{figure*}

In the left panel of Fig. \ref{fig:LinewidthsAmplitudes}, we plot the linewidths $\Gamma$ of the synthetic Rossby modes, as a function of $m$, for several values of the turbulent Reynolds number $\ret$. We also show, on the same plot, the observed linewidths reported by \citet{liang19} for solar Rossby modes at the equator. While the order of magnitude is consistent with the observations, the uncertainties associated with them do not permit us to discriminate between the different models. Of particular interest is the fact that we recover, for high $m$, and especially for the $\ret = 300$ case, the same $m^2$ dependence that can be inferred from the observations. This law is consistent with the theoretical Rossby mode dispersion relation (also shown on the plot), whose imaginary part yields
\begin{equation}
    \Gamma_R = \nut k_x^2~.
    \label{eq:RossbyTheoreticalLinewidth}
\end{equation}
We note that the value of the turbulent Reynolds number that seems to give the best agreement between the theoretical dispersion relation and the observations is $\ret = 300$, which corresponds to a turbulent viscosity $\nut = 570$ km$^2$ s$^{-1}$. This value is in accordance with the surface value inferred by \citet{gizon20}. However, it is significantly larger than the upper limit of $100$ km$^2$ s$^{-1}$ inferred by \citet{gizon21}, which was obtained under the assumption that the turbulent viscosity is constant over the entire convection zone.

Finally, we report the amplitude of the synthetic Rossby modes in the right panel of Fig. \ref{fig:LinewidthsAmplitudes}, in terms of latitudinal velocity $u_y$, for several models corresponding to different values of $\ret$. We also superimposed the observed amplitude reported by \citet{liang19}. The agreement that we find between the observations and the amplitudes yielded by our synthetic spectrum model, especially for $\ret = 700$ and $1000$, is consistent with our initial hypothesis that the inertial modes observed on the Sun are stochastically excited by turbulent convection. More specifically, we find that the amplitude of the equatorial Rossby modes initially increases with $m$, and then reaches a plateau where it remains fairly independent of $m$. The low amplitude of the low-m equatorial Rossby modes explains why the $m = 1$ or $2$ equatorial Rossby modes elude observation on the surface of the Sun. These results also show that increasing the turbulent viscosity (i.e. decreasing $\ret$) causes both an earlier start of the plateau and a lower value thereof. For instance, a value $\ret = 300$ causes the equatorial Rossby modes to reach an amplitude of $1.5$ m/s after $m \sim 10$, while a value $\ret = 1000$ causes them to reach an amplitude of $2.5$ m/s after $m \sim 15$.

\begin{figure}
    \centering
    \includegraphics[width=\linewidth]{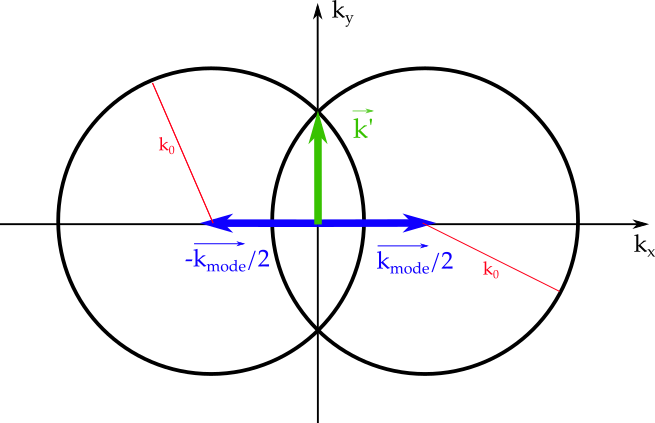}
    \caption{Geometrical argument used to identify the modes, $\mathbf{k}_{\rm mode}$, that can be excited by a single scale of turbulence, $k_0$. %In the simple case where the turbulent vorticity spectrum is characterised by a unique scale $k_0$, 
    The integrand in Eq. \ref{eq:FastVariableIntegral} peaks at $\mathbf{k}'$ such that both $|\mathbf{k}' + \mathbf{k}_\mathrm{mode}/2|$ and $|\mathbf{k}' - \mathbf{k}_\mathrm{mode}/2|$ are equal to $k_0$. The driving can therefore only be efficient if the two black circles in the figure intersect, i.e. if $k_\mathrm{mode} < 2 k_0$.}
    \label{fig:geometrical}
\end{figure}

The fact that the amplitudes predicted by our simple 2D model compare well with the solar observations deserves some discussion. This good agreement suggests that neither the radial eigenfunctions nor the source function depend strongly on radius $r$. Regarding the eigenfunctions, linear eigenmode computations in 3D by \citet{bekki22b} indicate that the dependence on $r$ is less pronounced for the smallest $m$ than for the larger $m$. For example, the velocity eigenfunction of the equatorial Rossby mode scales like the slowly varying function $r^m$ for $m = 3$ and $4$. Regarding the source of excitation, a turbulent vorticity spectrum that peaks at a scale $k_0$ can only drive modes with azimuthal wavenumbers $k_x = m/R < 2 k_0$ (see Fig.~\ref{fig:geometrical}). Since the radial vorticity for these large scales does not vary fast with depth \citep[see e.g.][their Fig.~13e]{miesch08}, it is perhaps not too surprising that our 2D model predicts the amplitudes of the low-$m$ inertial modes correctly (in order of magnitude). Future work should nevertheless account for the radial dependence of the source properties, as well as the mode eigenfunctions and viscous damping.

\subsubsection{Visibility of the modes\label{subsubsec:ComparisonObs}}

So far, we have only investigated the component of the spectrum caused by the resonant inertial modes, and we have excluded the turbulent noise component from our analysis. However, the inclusion of this noise component is necessary in order to assess whether the signal of the modes is visible above the noise level. To do this, we simply add the $u_y$ noise component given by Eq. \ref{eq:NoiseUy} to the inertial mode component given by Eq. \ref{eq:PowerSpectrumUy}. Furthermore, we would like to directly compare our results to the observed equatorial $u_\theta$ spectrum reported by \citet{liang19}. The authors measured south-north helioseismic travel times along the solar equator using datasets from the Michelson Doppler Imager (MDI) on board the Solar and Heliospheric Observatory (SoHO) and the Helioseismic and Magnetic Imager (HMI) on the Solar Dynamics Observatory (SDO). Their procedure was carried out for several latitudes uniformly distributed within the range $\lambda = \pm 15^\circ$. Therefore, to mimic their procedure, we average our synthetic $u_y$ power spectrum over this latitude range (that is, we average between $\yo = \pm \sin(15^\circ)$ instead of simply taking $\yo = 0$).

\begin{figure}
    \centering
    \begin{tabular}{c}
        \includegraphics[width=\linewidth]{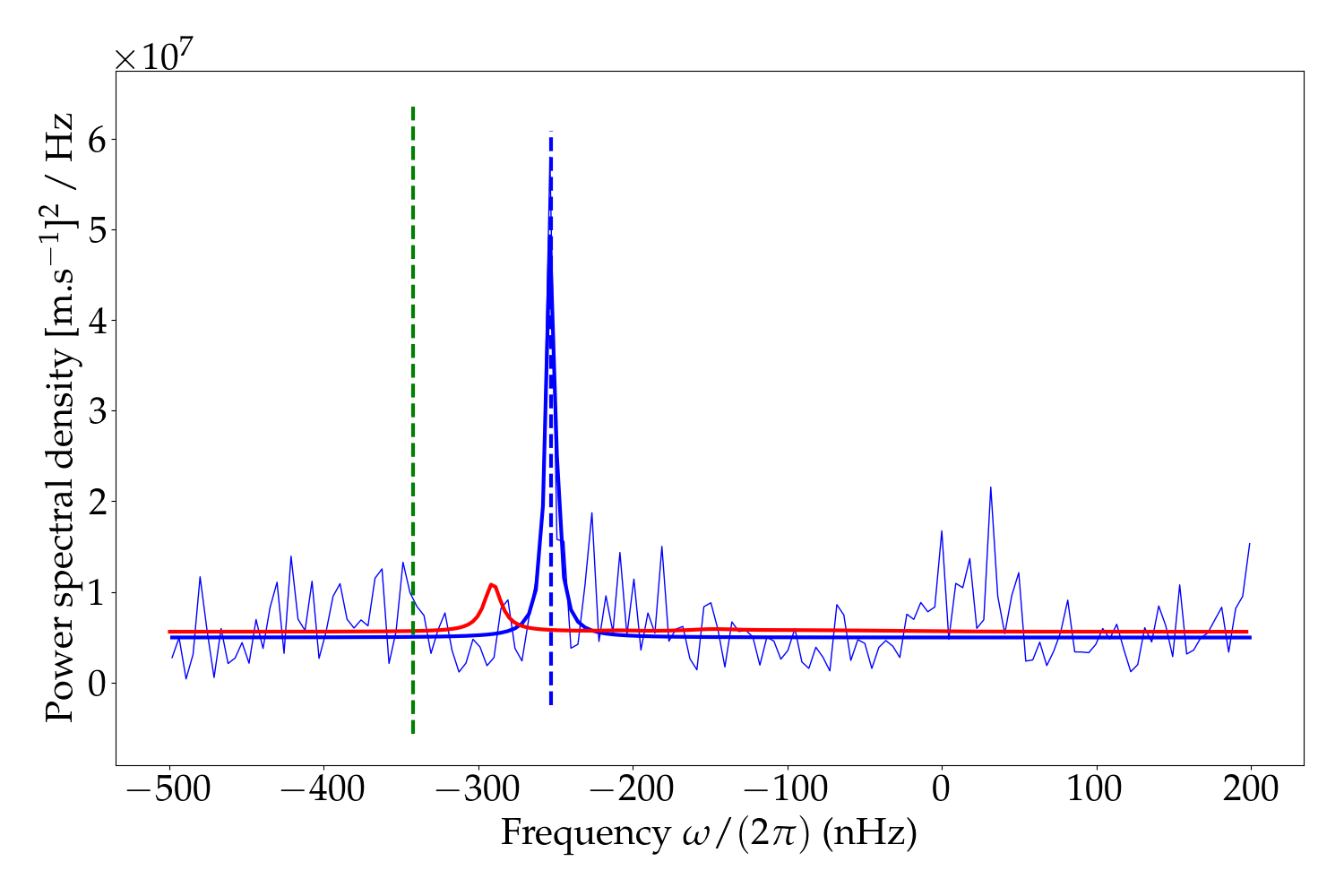} \\
        \includegraphics[width=\linewidth]{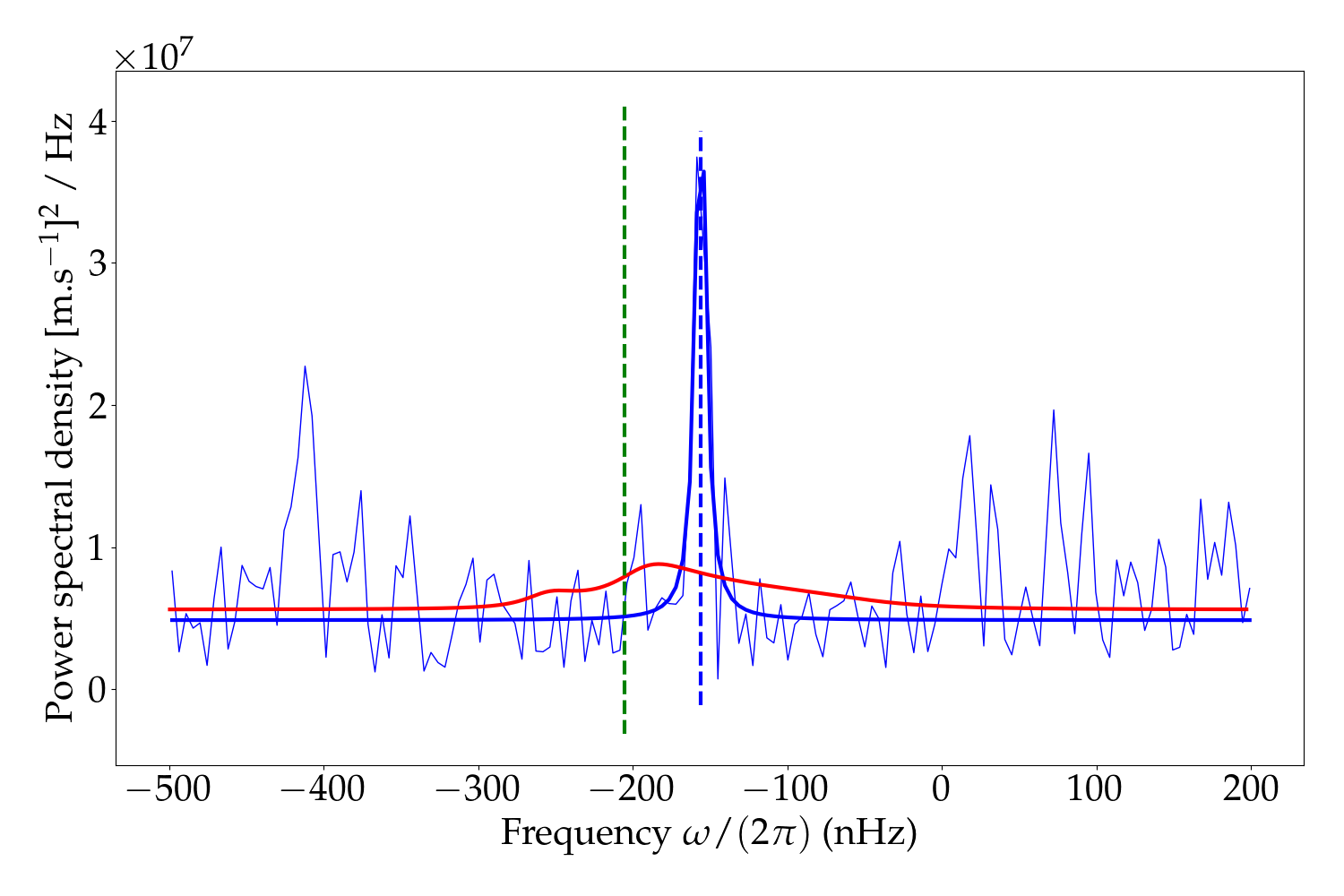} \\
        \includegraphics[width=\linewidth]{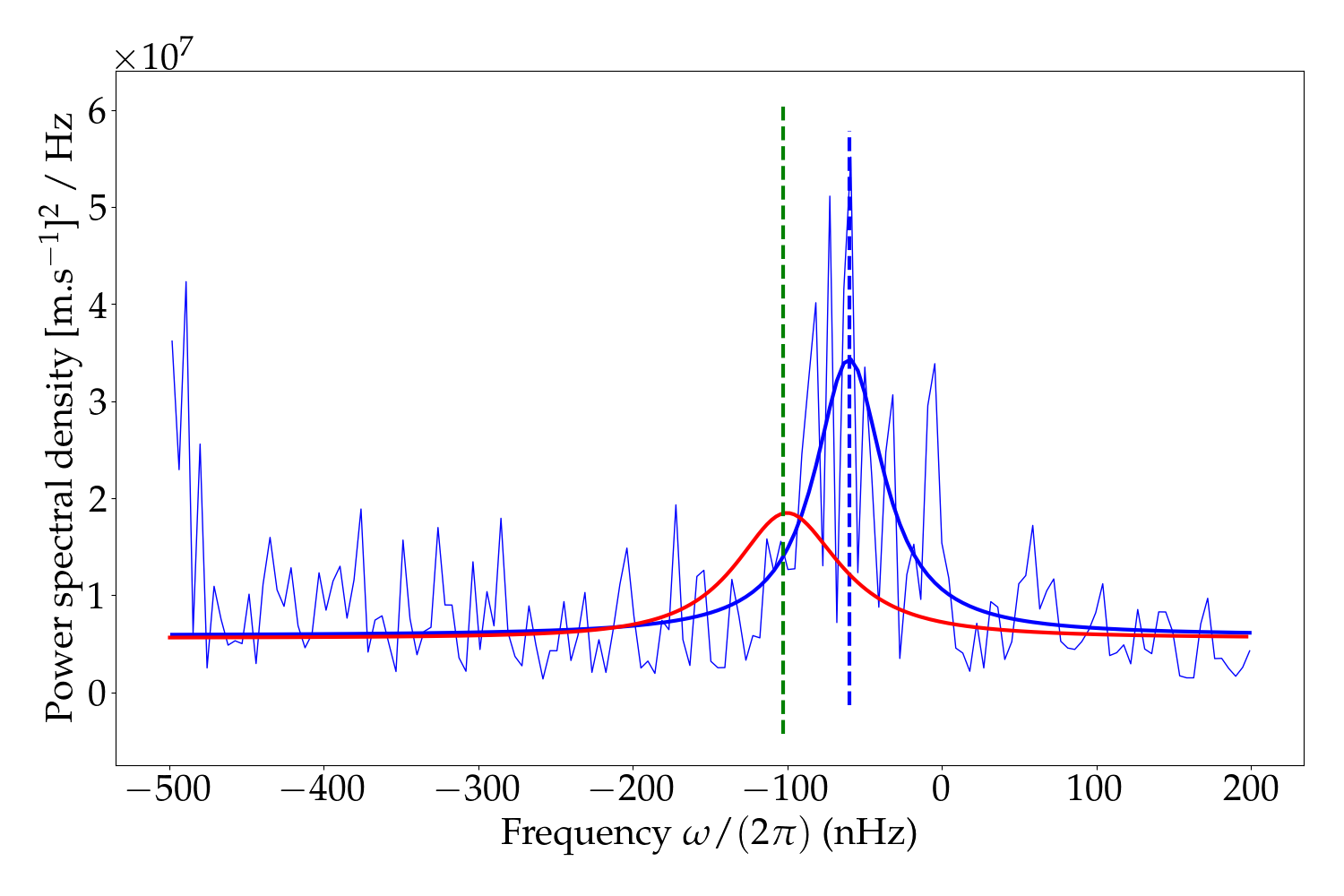}
    \end{tabular}
    \caption{Latitudinal velocity spectrum, estimated at the solar equator, for azimuthal orders $m = 3$ (\textbf{top}), $5$ (\textbf{middle}), and $10$ (\textbf{bottom}), and for a turbulent Reynolds number $\ret = 300$. The solid red line represents our model, including the contribution both from the inertial modes and from the turbulent noise. The thin blue line shows the solar observations reported by \citet{liang19}, and the thicker blue line shows the best Lorentzian fit to their data, as reported in their paper.}
    \label{fig:ObsVsTheory}
\end{figure}

The comparison is shown in Fig. \ref{fig:ObsVsTheory} for azimuthal orders $m = 3$, $5$ and $10$. It can be seen that, by construction, the turbulent noise level in our synthetic power spectrum matches the noise level in the corresponding solar observations. We also point out, before diving further in the question of mode visibility, that the central frequencies in our model do not correspond to the observed mode frequencies, especially for higher values of $m$. This is to be expected, because the homogeneous part of the wave equation governing the frequencies of the inertial modes is somewhat simplified in our model (in particular the fact that we reduced the system to a 2D model, or the fact that we did not account for the spherical geometry of the domain). The focus of this study is not on the mode frequencies, so that these discrepancies do not constitute an issue.

We find that for $m = 1$ and $m = 2$, the amplitude of the modes is lower than the turbulent noise level. On this, our model prediction agrees with observational evidence; \citet{liang19} had indeed reported that no significant peak could be detected for such low values of the azimuthal order.

For $m = 3$ to $5$, we recall that several peaks could be distinguished in our synthetic spectrum. However, as illustrated for $m = 3$ in the top panel of Fig. \ref{fig:ObsVsTheory}, it turns out that only one of these resonant peaks has enough amplitude to rise above the turbulent noise level. This is also in agreement with the observations.

For $m \geqslant 5$, we recall that only one peak is predicted to dominate in the equatorial spectrum, corresponding to the equatorial Rossby mode. At the start of this $m \geqslant 5$ regime, this mode is clearly visible above the noise level (as illustrated for $m = 10$ in the bottom panel of Fig. \ref{fig:ObsVsTheory}). However, we find that the visibility of the mode becomes more and more questionable as $m$ increases. This is in agreement with the solar observations reported by \citet{liang19}, who could not detect any significant peak in the equatorial spectrum for $m \geqslant 16$.

Our model allows some light to be shed on this high-m visibility issue. We show in Sect. \ref{subsubsec:AmplitudesLinewidths} that, in this high-m regime, the amplitudes of the equatorial Rossby modes are fairly independent of $m$, while their linewidths increase as $m^2$. Because the peaks become wider and wider for the same total power, their spectral height decreases as $1/m^2$, thus explaining why they end up below the turbulent noise level after some point. We point out that, because the modes are stochastically excited by turbulent convection, their signal-to-noise ratio is inherent to the excitation process, and not contingent on the observational setup. It would therefore not be possible to improve their visibility through a longer observation time period.

\subsection{Power spectrum in the frequency--latitude plane\label{subsec:2DSpectra}}

\begin{figure}
    \centering
    \begin{tabular}{c}
        \includegraphics[width=\linewidth]{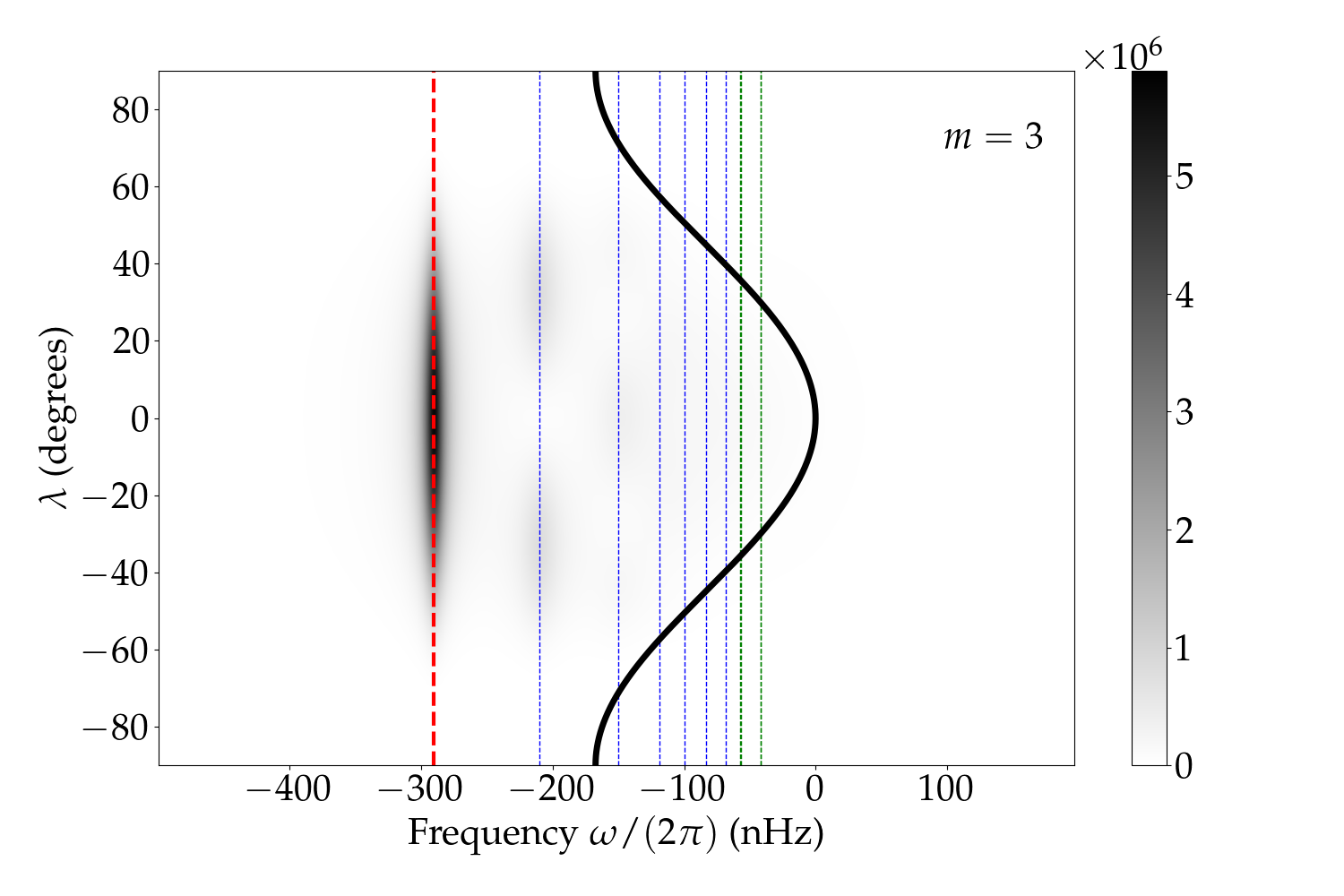} \\
        \includegraphics[width=\linewidth]{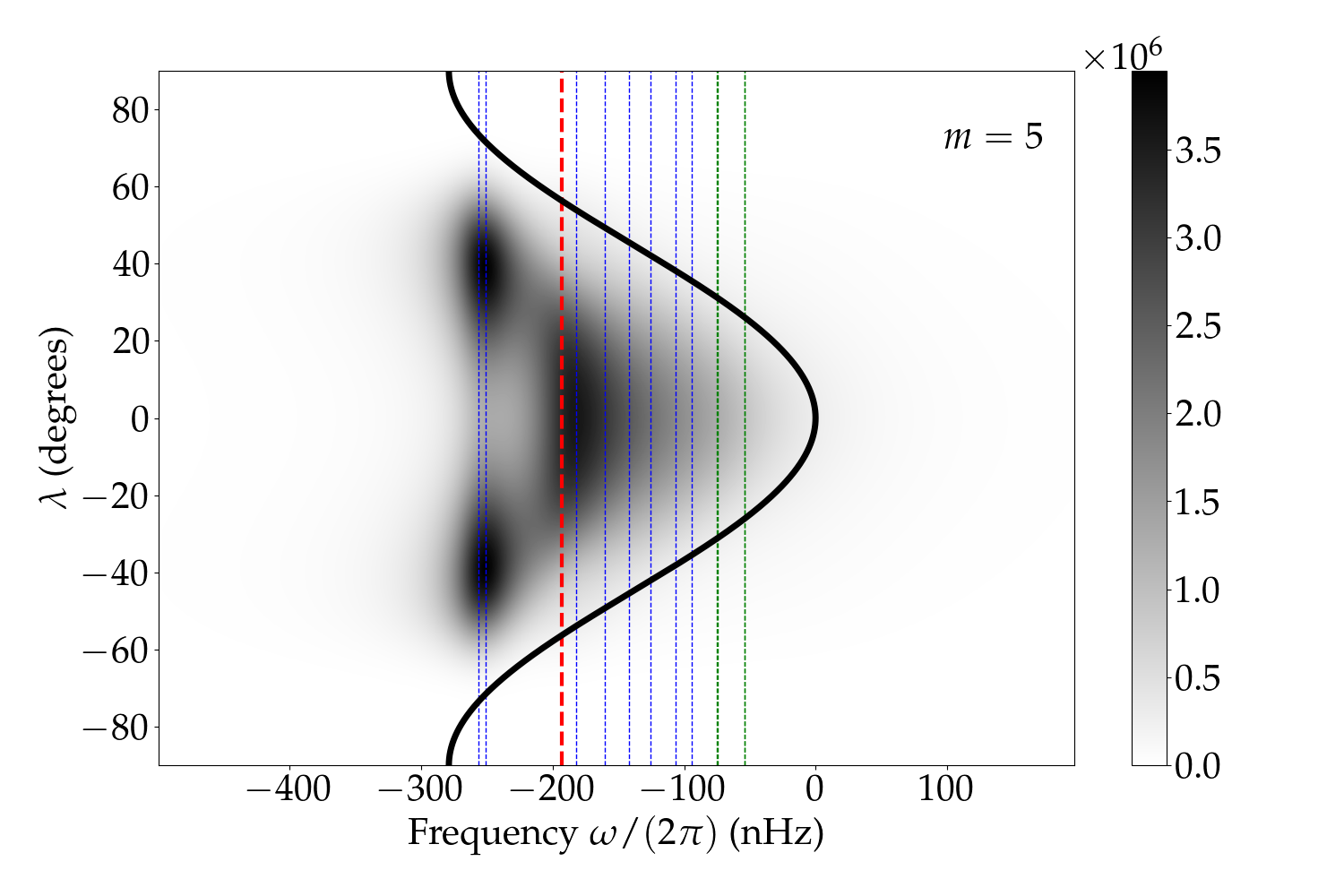} \\
        \includegraphics[width=\linewidth]{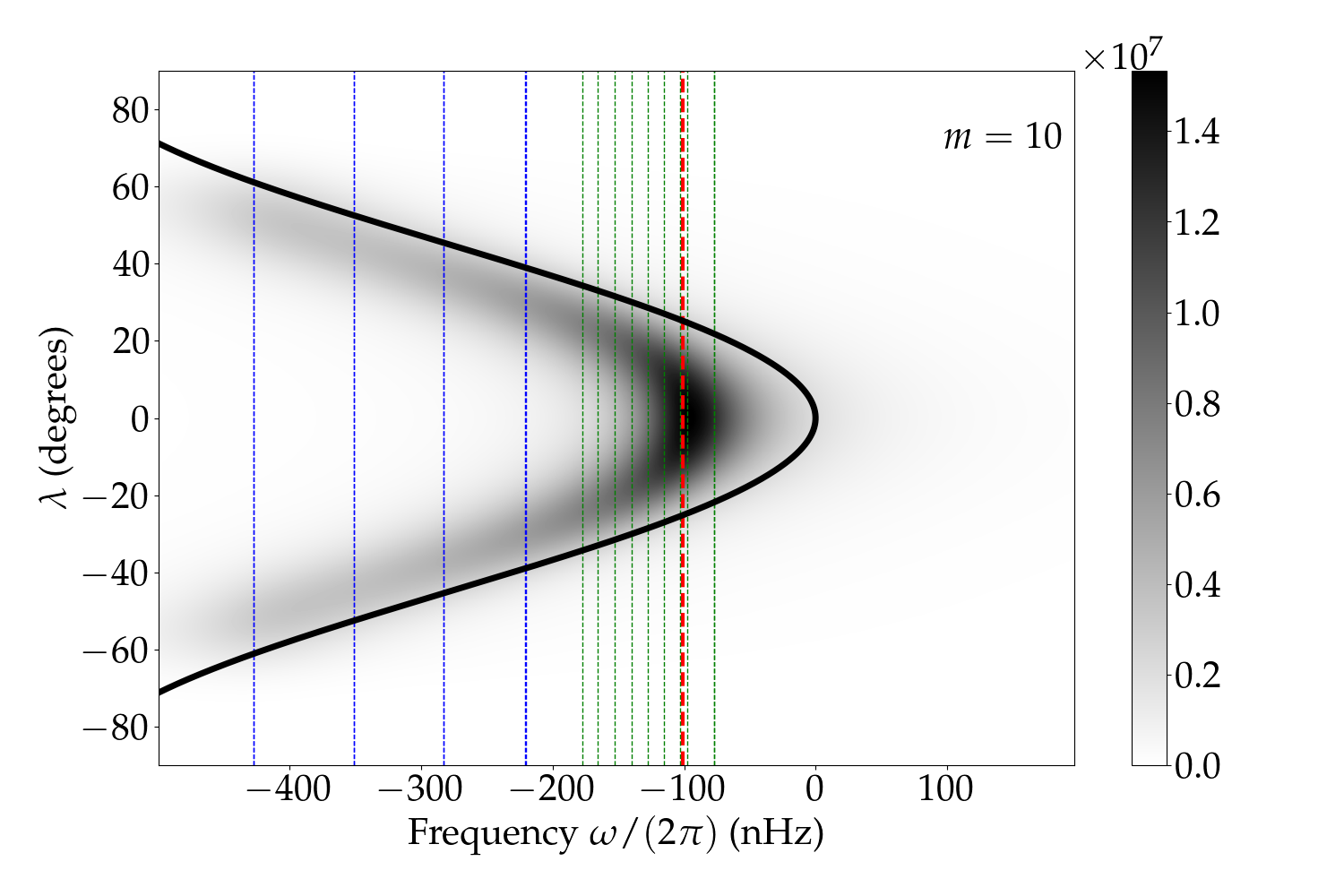}
    \end{tabular}
    \caption{Power spectrum of the latitudinal velocity, $u_y$, as a function of frequency (horizontal axis) and latitude (vertical axis). Each panel corresponds to a different azimuthal order: $m = 3$ (\textbf{top}), $m = 5$ (\textbf{middle}), and $m = 10$ (\textbf{bottom}). The turbulent Reynolds number is set to $\ret = 300$. The solid black curve shows the critical latitudes, where the differential rotation exactly matches the phase velocity of the inertial waves, and is defined by the implicit relation Eq. \ref{eq:CriticalLatitudes}. The dashed vertical lines show the real part of the eigenfrequencies of the homogeneous problem, with the same colour code as in Fig. \ref{fig:Frequencies}.}
    \label{fig:2DSpectraTheta}
\end{figure}

Throughout Sect. \ref{subsec:EquatorialSpectra}, we focused our analysis on equatorial power spectra. However, our model also gives us access to the power spectra at any given latitude. This is illustrated in Fig. \ref{fig:2DSpectraTheta}, for the $u_y$ power spectrum and for azimuthal orders $m = 3$, $5,$ and $10$. The solid black curve in each panel corresponds to the critical layer where the azimuthal velocity, $U$, stemming from differential rotation exactly matches the azimuthal phase velocity of the inertial waves, that is, the curve $y_c(\omega)$ given by the following implicit relation:
\begin{equation}
    U(y_c) = \dfrac{\omega}{k_x} = \dfrac{\omega R}{m}~,
    \label{eq:CriticalLatitudes}
\end{equation}
where $U$ is given by Eq. \ref{eq:DifferentialRotation}.

The shape of the power spectrum clearly transitions from a low-m regime, dominated by a few, clearly identifiable and distinguishable resonant modes, to a high-m regime, where the power is concentrated in a characteristic crescent-shaped region along the critical latitude (or, more precisely, just below the critical latitude). The transition between the two regimes occurs at $m \sim 5$ and corresponds to the same transition that we already mentioned for the equatorial spectra in Sect. \ref{subsec:EquatorialSpectra}. This seems to indicate that the detection of inertial modes in observational data is much more delicate for high $m$ than for low $m$: the overlapping of the excess power regions associated with each mode in the frequency -- latitude plane is indeed likely to make the interpretation of the observed spectra considerably more complex.

\begin{figure}
    \centering
    \begin{tabular}{c}
        \includegraphics[width=\linewidth]{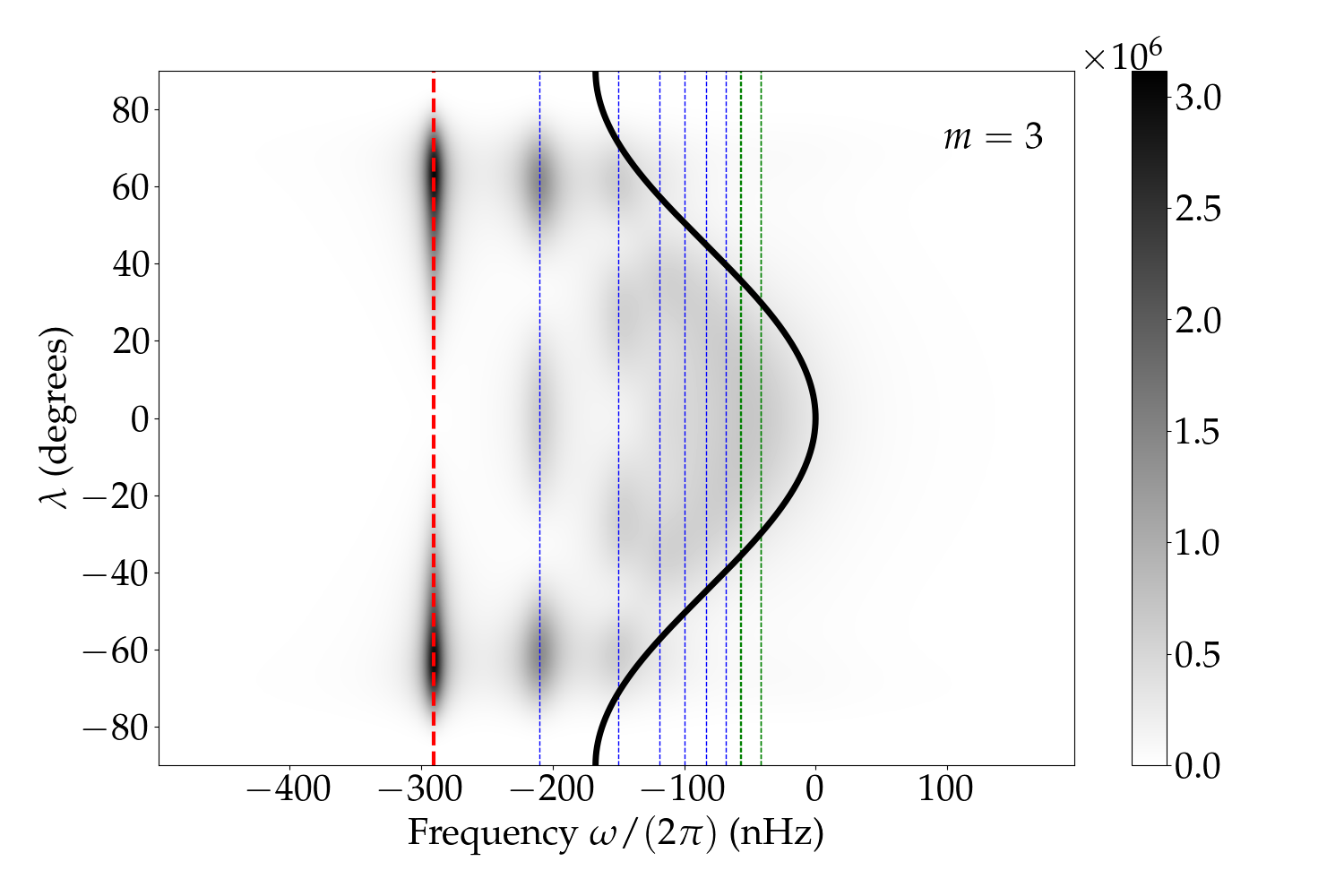} \\
        \includegraphics[width=\linewidth]{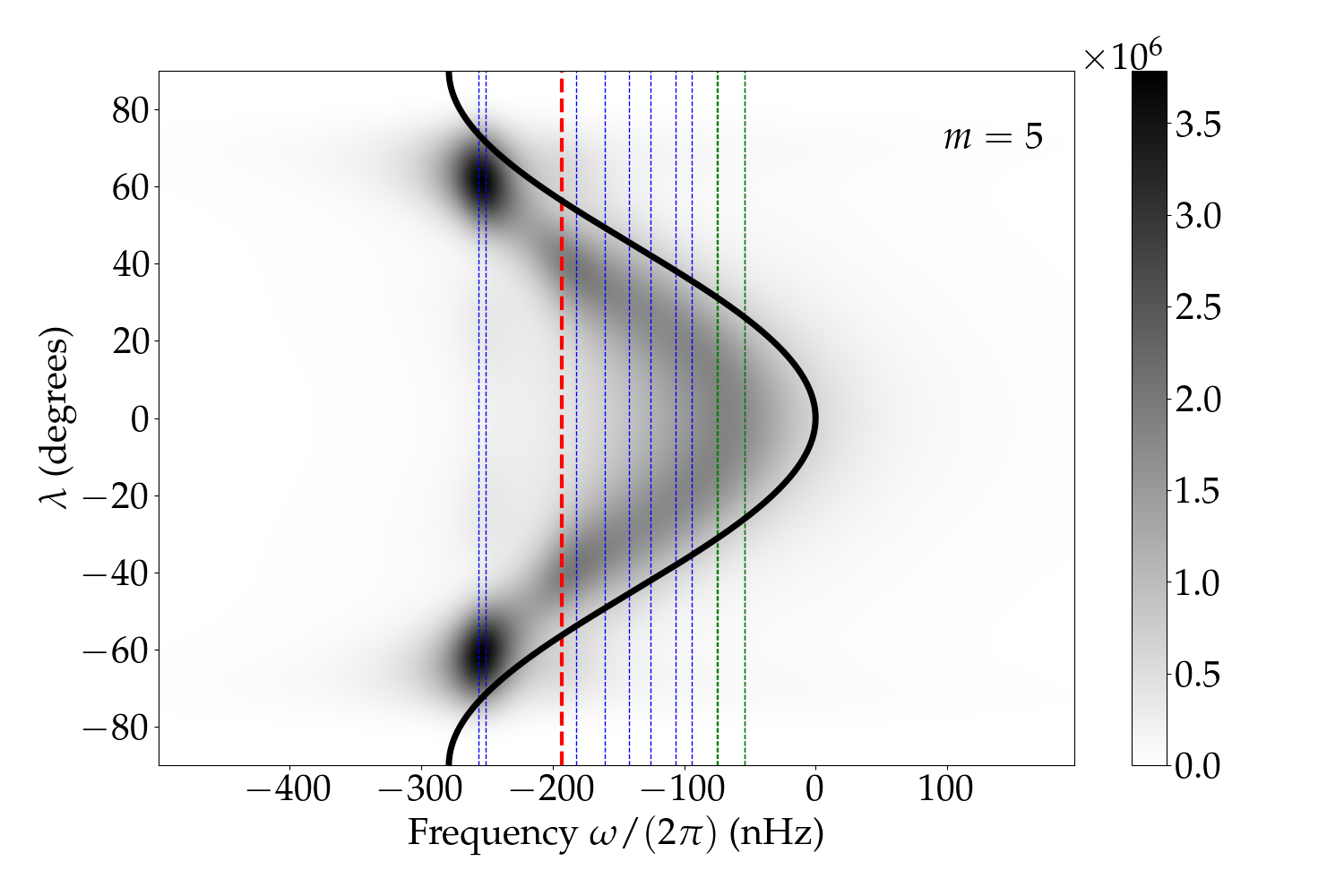} \\
        \includegraphics[width=\linewidth]{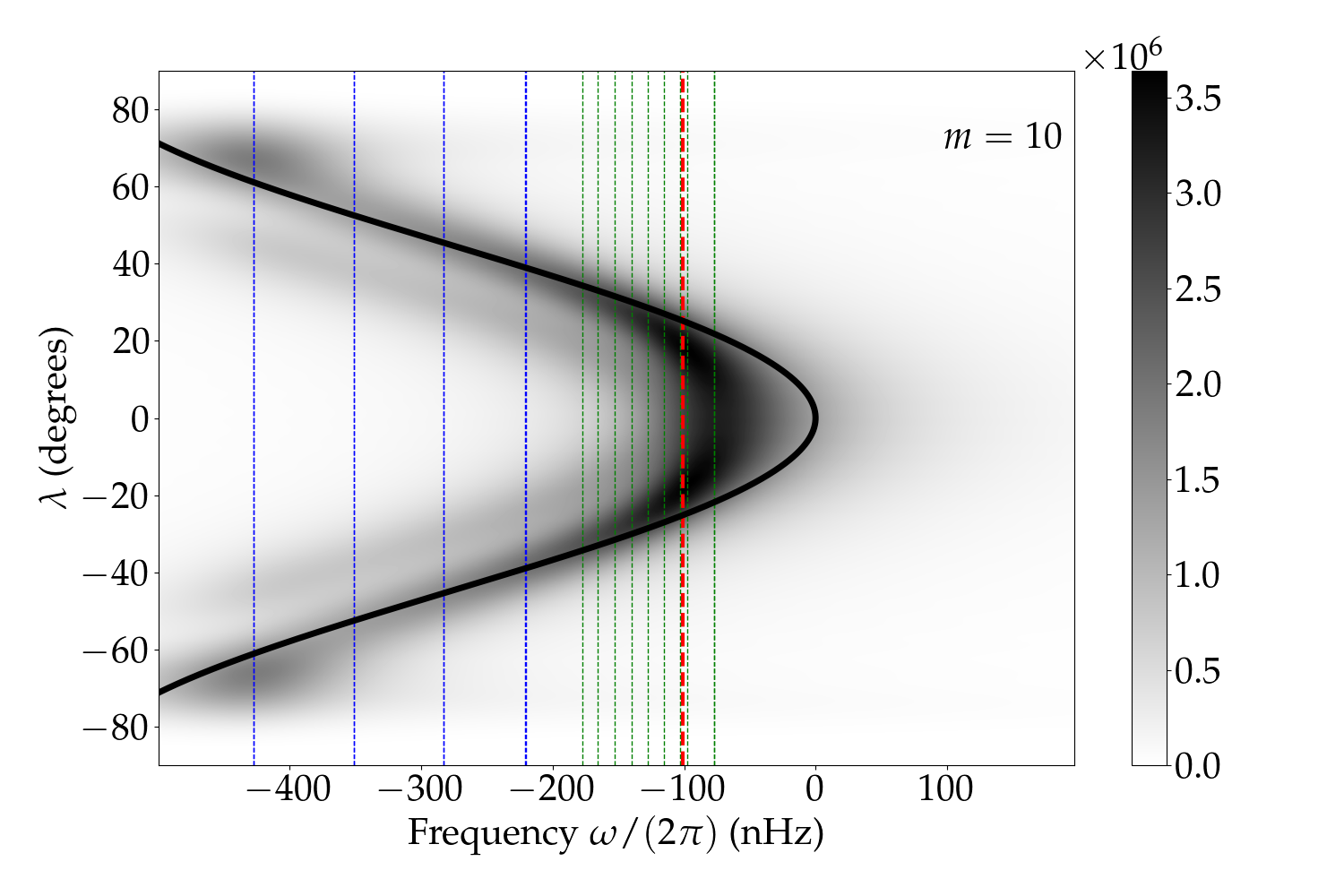}
    \end{tabular}
    \caption{Same as Fig. \ref{fig:2DSpectraTheta}, but for the power spectrum of the azimuthal velocity, $u_x$.}
    \label{fig:2DSpectraPhi}
\end{figure}

Similarly, Fig. \ref{fig:2DSpectraPhi} shows the $u_x$ power spectrum in the same frequency -- latitude plane. While the same low-m to high-m regime transition can be observed, there is still a number of qualitative differences with the $u_y$ power spectrum. The main difference is that, for high values of $m$, the region where most of the power is located passes across the critical latitude as frequency becomes more and more negative, while this region was kept confined at lower latitudes in the $u_y$ spectrum. This is simply a symptom of the qualitative difference between the $u_y$ and $u_x$ eigenfunctions. The effect grows stronger as $m$ increases, and indicates that, depending on the observable used to detect inertial mode, the region just above the critical latitude may be of similar interest than the region below.

\begin{figure}
    \centering
    \begin{tabular}{c}
        \includegraphics[width=\linewidth]{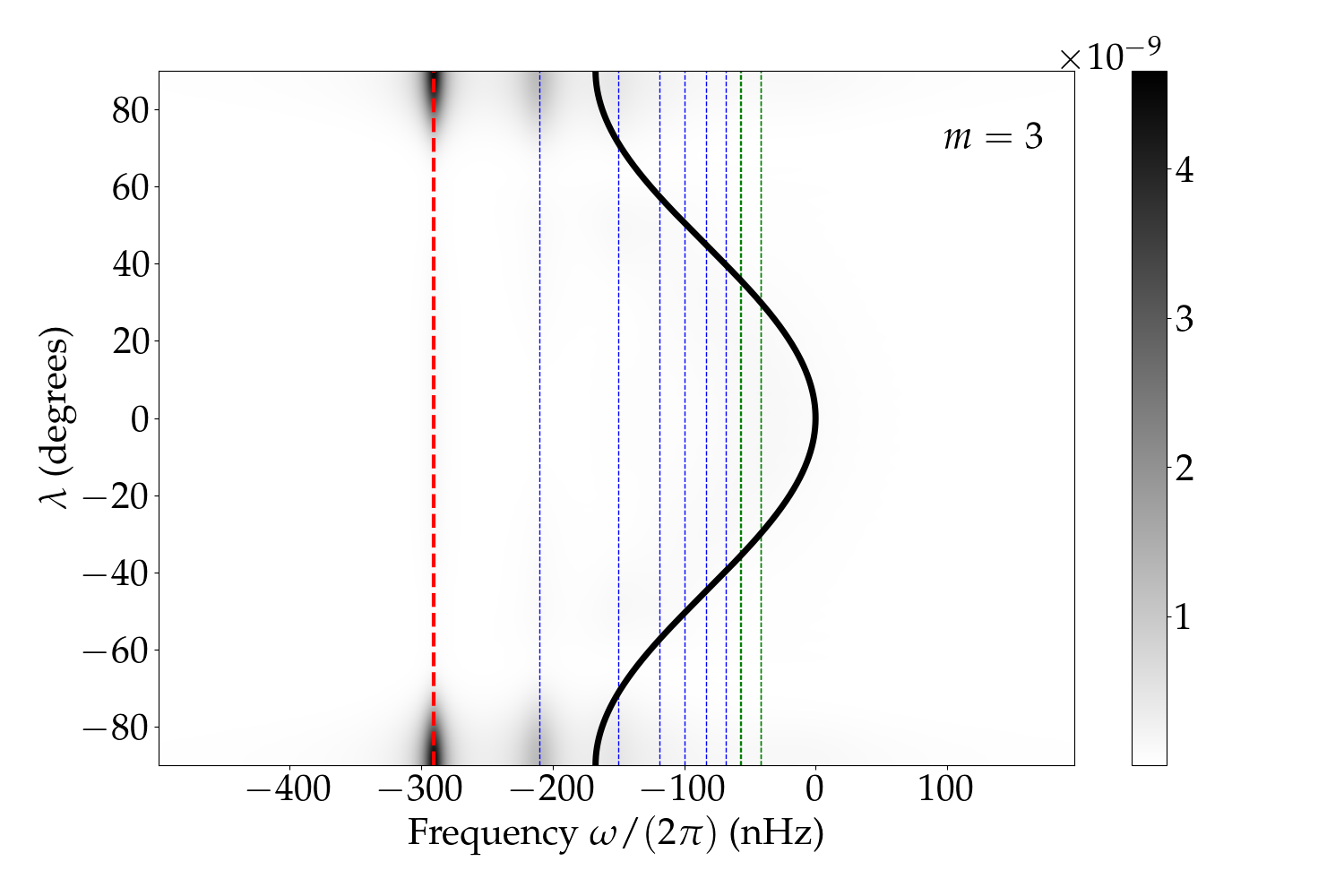} \\
        \includegraphics[width=\linewidth]{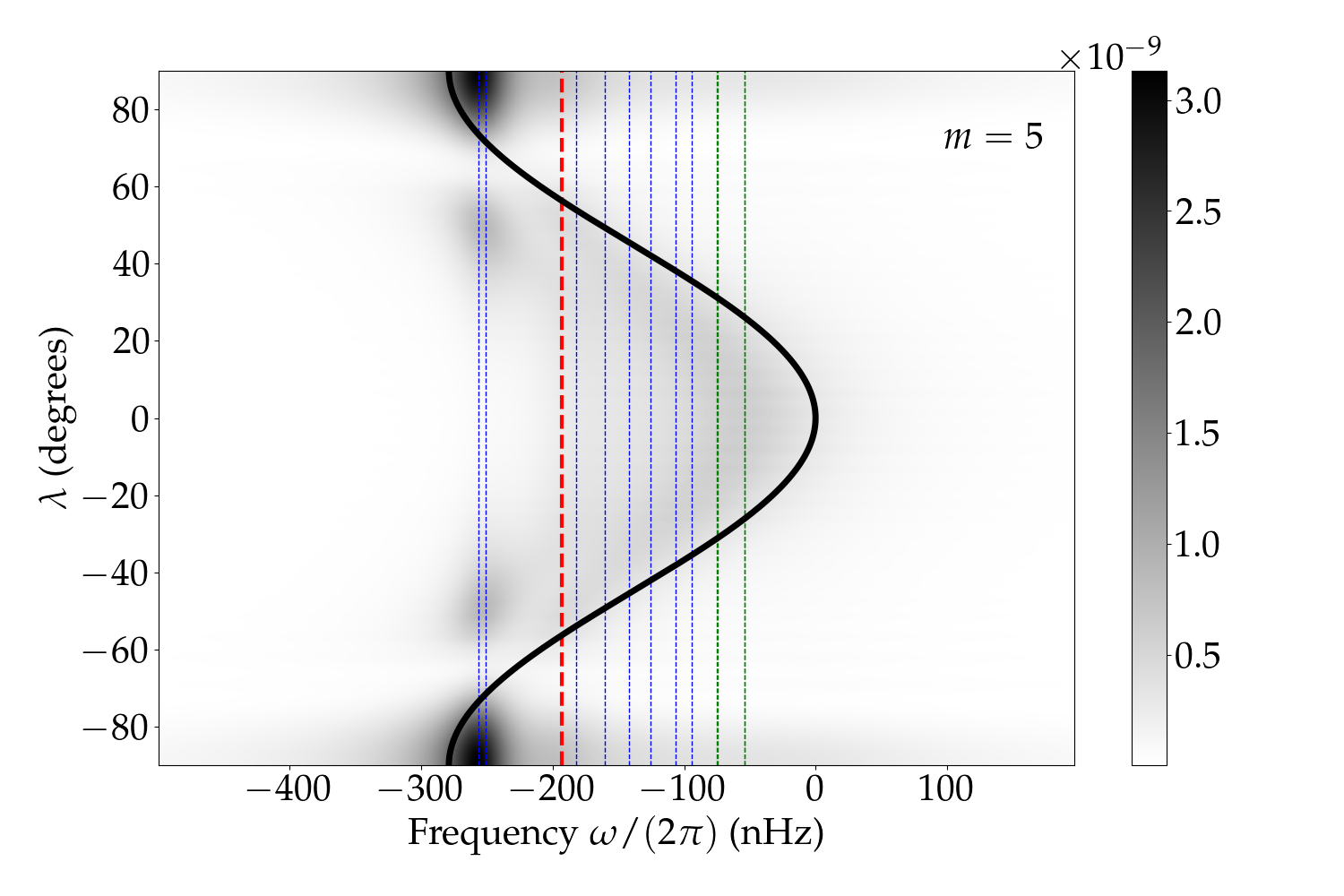} \\
        \includegraphics[width=\linewidth]{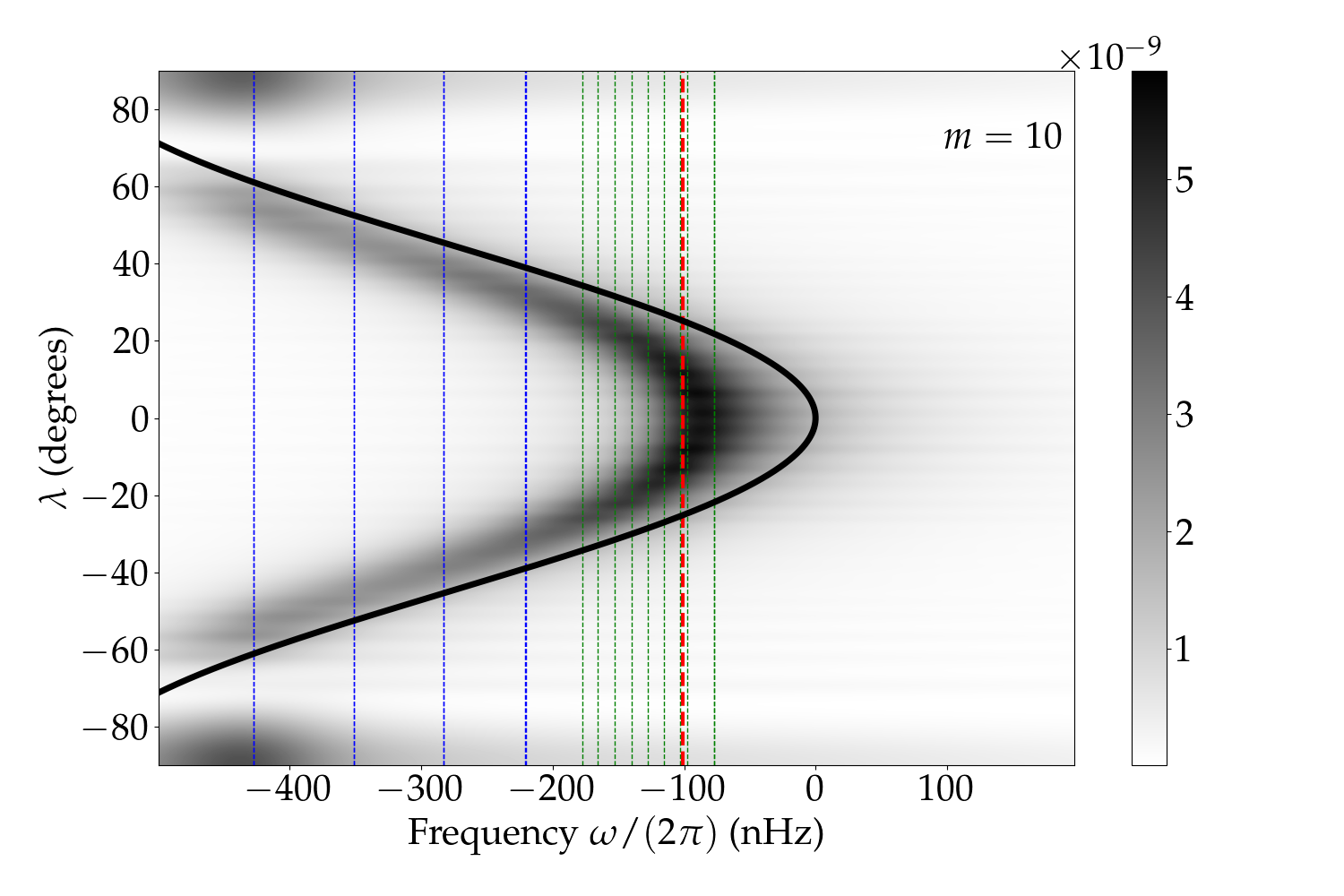}
    \end{tabular}
    \caption{Same as Fig. \ref{fig:2DSpectraTheta}, but for the power spectrum of the radial vorticity, $\zeta$.}
    \label{fig:2DSpectraVort}
\end{figure}

We show similar results for the vorticity power spectrum in Fig. \ref{fig:2DSpectraVort}. Again, the main difference comes from the latitudinal structure of the eigenfunctions, which contains significantly more power at the poles than the $u_y$ or $u_x$ eigenfunctions. However, the low-m to high-m transition described above can still be observed in the vorticity power spectra.

%%% Conclusion

\section{Conclusion\label{sec:conclusion}}

We have designed a model for the stochastic excitation of linearly stable, quasi-toroidal solar inertial modes by turbulent convection. In order to do so, we adopted a simplified 2D framework, where inertial modes are described in an equatorial $\beta$ plane close to the surface of the Sun. We included latitudinal differential rotation in the form of a parabolic, Poiseuille-like profile, with values chosen to best approximate the solar differential rotation at low latitudes. Using this model, we successfully reproduce the observed amplitude of the linearly stable, low- and mid-latitude inertial modes, with latitudinal velocities ranging between $\sim 0.1$ and $\sim 1.5$ m.s$^{-1}$, similar to those reported, for instance, by \citet{liang19}. The amplitude of the linearly stable inertial modes observed in the equatorial region of the Sun is therefore consistent with a stochastic excitation by turbulent convection. However, we did not treat the case of the unstable, high-latitude inertial modes.

We also show that the power spectra in the frequency--latitude plane have a very different qualitative behaviour depending on whether $m \lesssim 5$ or $m \gtrsim 5$. In the low-$m$ regime, the spectra are dominated by non-overlapping, clearly identifiable and distinguishable resonant modes. On the other hand, in the high-$m$ regime, the line profile of the modes is much wider and so the excess power region associated with each mode overlap, thus forming a single crescent-shaped excess power region along the critical latitude. In this regime, it is much harder to distinguish between individual modes. This has important implications for the identification of inertial modes in solar data, as this seems to indicate that the interpretation of the observed spectra becomes increasingly complex as $m$ increases.

In the equatorial spectra, the predicted amplitudes of the modes are such that they are only visible above the turbulent noise level for $m \geqslant 3$, in accordance with solar observations. Between $m = 3$ and $5$, the equatorial spectra feature several dominant peaks, whose line profiles do not overlap. By contrast, for $m > 5$, the power spectrum is dominated by the inertial modes with the longest lifetime, and which correspond to the Cartesian equivalent of the classical equatorial Rossby modes. We predict that the amplitude of these equatorial Rossby modes will increase with $m$ until $m \sim 10$, after which the amplitudes reach a plateau and become fairly independent of azimuthal order. By contrast, their linewidth increases with $m$ such that their spectral height decreases in such a way that the mode stops being visible above the turbulent noise level for $m \sim 16$, in accordance with observations. Interestingly, we find that some modes incur mutually destructive interference, to such a degree that their overall amplitude is negligible even when their individual amplitudes should make them visible. This is possible because all the modes share the same source of excitation.

Additionally, we find that the theoretically predicted full linewidths at half maximum of the equatorial Rossby modes agree reasonably well with the observed linewidths, provided we choose the turbulent viscosity to be $\sim 570$ km.s$^{-1}$. This confirms constraints previously obtained in the literature for the convection-induced turbulent viscosity. We also show that the linewidths of the equatorial Rossby modes vary as $m^2$, which can be predicted through the classical Rossby wave dispersion relation (see Eq. \ref{eq:RossbyTheoreticalLinewidth}). Because this simple square law primarily depends on the turbulent viscosity,  the value of the turbulent viscosity can be constrained throughout the solar convection zone, even potentially through the use of inversion techniques.

While the equatorial $\beta$-plane approximation constitutes a drastic simplification for low azimuthal orders, $m$, it is not so much the case for higher $m$. Therefore, we do not expect our results to be overly affected should this approximation be lifted, and should the derivations be carried out in spherical geometry. This would nevertheless warrant further investigation. The suppression of the radial coordinate in the problem, by contrast, undoubtedly constitutes a more important approximation. In particular, the use of a 3D model, rather than 2D, would increase the density of modes in the eigenspectrum, and therefore the complexity of the predicted power spectrum. We find that the present 2D model makes relevant predictions in the case of quasi-toroidal modes; however, going from 2D to 3D remains necessary to obtain more accurate mode amplitudes and to constrain the radial dependence of the turbulent viscosity and the source properties.

The present synthetic spectrum model may also be of interest to test mode detection pipelines. To that effect, specific spectrum realisations need to be drawn from the expected power distribution. This not only requires knowledge of the expected power spectrum (i.e. the variance associated with the complex Fourier transform for each pixel in the frequency--latitude plane), but also the correlation matrix between the different frequencies and latitudes. While different frequencies are completely uncorrelated under the hypothesis that the source of excitation is a stationary stochastic process, the correlations between different latitudes must be accounted for. The present framework would allow us to do this with only minimal adaptation. This constitutes one of the uses to which the present model will be put in the near future.

Finally, as we pointed out before, the present model is not meant to treat the case of unstable inertial modes. Some high-latitude modes can be shown to be self-excited, either because of a steep rotational shear  \citep{fournier22} or by a baroclinic instability \citep{bekki22b}.  The latter is responsible for the unstable nature of the $m = 1$ high-latitude mode, which is easily  identifiable in the solar observations. The amplitude reached by this mode is related to a non-linear saturation process, which is well outside the scope of the present linear study.

%%% Acknowledgements

\begin{acknowledgements}
We are very grateful to  Zhi-Chao Liang for providing the Rossby-mode observations and Christian Baumgartner for the surface vorticity data. This work is supported in part by the ERC Synergy Grant WHOLESUN 810218.
\end{acknowledgements}

%%% References

\bibliographystyle{aa}
\bibliography{AA_2022_45666}

%%% Appendices

\begin{appendix}

%%% Appendix A

\section{Vorticity wave equation in the equatorial $\beta$ plane\label{app:WaveEquation}}

\subsection{Transport equation for the stream function}

The total flow velocity $\mathrm{\mathbf{v}}$ is decomposed into a background zonal flow $\mathbf{U} \equiv U(y) ~ \mathbf{e_x}$, which represents differential rotation, and a residual flow $\mathbf{u} \equiv u_x(x,y) \mathbf{e_x} + u_y(x,y) \mathbf{e_y}$, which contains both the waves and the convective noise. The equations of motion become
\begin{align}
    & \dfrac{\partial u_x}{\partial t} + U\dfrac{\partial u_x}{\partial x} + u_y U' + u_x \dfrac{\partial u_x}{\partial x} + u_y \dfrac{\partial u_x}{\partial y} = -\dfrac{1}{\rho} \dfrac{\partial p}{\partial x} + f u_y~, \\
    & \dfrac{\partial u_y}{\partial t} + U\dfrac{\partial u_y}{\partial x} + u_x \dfrac{\partial u_y}{\partial x} + u_y \dfrac{\partial u_y}{\partial y} = -\dfrac{1}{\rho} \dfrac{\partial p}{\partial y} - f u_x~ ,
\end{align}
where $U'=dU/dy$.
Assuming that the residual flow is  incompressible, there exists a scalar field $\Psi$,  the stream function, such that
\begin{equation}
    \mathbf{u} = \bm{\nabla} \wedge \left( \Psi \mathbf{e}_z \right) ,
\end{equation}
where $\mathbf{e}_z$ is the radial unit vector. Explicitly,
\begin{align}
    & u_x = \dfrac{\partial \Psi}{\partial y}~, \\
    & u_y = -\dfrac{\partial \Psi}{\partial x}~.
\end{align}
Differentiating the $x$-component of the equation of motion with respect to $y$ and the $y$-component  with respect to $x$, and subtracting the two, one obtains the following equation:
\begin{multline}
    \dfrac{\partial \Delta\Psi}{\partial t} + U\dfrac{\partial \Delta\Psi}{\partial x} - U'' \dfrac{\partial \Psi}{\partial x} + \dfrac{\partial \Psi}{\partial y} \dfrac{\partial \Delta\Psi}{\partial x} - \dfrac{\partial \Psi}{\partial x} \dfrac{\partial \Delta\Psi}{\partial y} \\
    = \dfrac{1}{\rho^2}\left( \dfrac{\partial \rho}{\partial y} \dfrac{\partial p}{\partial x} - \dfrac{\partial \rho}{\partial x} \dfrac{\partial p}{\partial y} \right) - \beta \dfrac{\partial \Psi}{\partial x}~,
\end{multline}
where $\Delta \equiv \partial_x^2 + \partial_y^2$ is the horizontal Laplacian operator. In the case of a non-adiabatic flow, the first term on the right-hand side does not vanish, because the density and pressure streamlines are distinct. However, if we disregard non-adiabatic effects, the flow can be considered barotropic, meaning that the gas pressure is a function of density only,
\begin{equation}
    p = p(\rho)~.
\end{equation}
Then the gas pressure gradient and density gradient are aligned, and we have
\begin{equation}
    \dfrac{\partial \rho}{\partial y} \dfrac{\partial p}{\partial x} - \dfrac{\partial \rho}{\partial x} \dfrac{\partial p}{\partial y} = (\bm{\nabla} p \wedge \bm{\nabla}\rho ) \cdot \mathbf{e}_z = 0~.
\end{equation}
We thus obtain a purely mechanical equation,
\begin{equation}
    \left( \dfrac{\partial}{\partial t} + U\dfrac{\partial}{\partial x}\right) \Delta\Psi + \left(\beta - U''\right) \dfrac{\partial\Psi}{\partial x} + \dfrac{\partial \Psi}{\partial y} \dfrac{\partial \Delta\Psi}{\partial x} - \dfrac{\partial \Psi}{\partial x} \dfrac{\partial \Delta\Psi}{\partial y} = 0~.
    \label{eqapp:StreamFunctionBeforeNut}
\end{equation}

\subsection{Turbulent viscosity}

The last two terms of Eq. \ref{eqapp:StreamFunctionBeforeNut} can be split into an average quantity and a fluctuation around this average. As we will see, the former gives rise to a turbulent viscous term in the wave equation, whereas the latter will act as a source term. We define

\begin{multline}
    \delta\left( \dfrac{\partial \Psi}{\partial y} \dfrac{\partial \Delta \Psi}{\partial x} - \dfrac{\partial \Psi}{\partial x} \dfrac{\partial \Delta \Psi}{\partial y} \right) \equiv \dfrac{\partial \Psi}{\partial y} \dfrac{\partial \Delta \Psi}{\partial x} - \dfrac{\partial \Psi}{\partial x} \dfrac{\partial \Delta \Psi}{\partial y} \\
    - \left\langle \dfrac{\partial \Psi}{\partial y} \dfrac{\partial \Delta \Psi}{\partial x} - \dfrac{\partial \Psi}{\partial x} \dfrac{\partial \Delta \Psi}{\partial y} \right\rangle_h~,
    \label{eqapp:DefinitionDelta}
\end{multline}
where the notation $\langle . \rangle_h$ refers to a horizontal average over scales larger than the turbulent scale, but smaller than the scale of the inertial modes. The scale separation that allows for the definition of this average is discussed in the main text (see Sect. \ref{subsec:TurbulentSpectrum}).

The average term can be rewritten as
\begin{equation}
    \left\langle \dfrac{\partial \Psi}{\partial y} \dfrac{\partial \Delta \Psi}{\partial x} - \dfrac{\partial \Psi}{\partial x} \dfrac{\partial \Delta \Psi}{\partial y} \right\rangle_h = \left\langle \vphantom{a_j^I} (\mathbf{u} \cdot \bm{\nabla}) \Delta\Psi \right\rangle_h = \bm{\nabla} \cdot \langle \mathbf{u} ~ \Delta \Psi \rangle_h~,
\end{equation}
where the second equality stems from the assumed incompressible nature of the flow, which translates to $\bm{\nabla} \cdot \mathbf{u} = 0$. It can be seen that this is the divergence of a mean flux representing the transport of the quantity $\Delta\Psi$ (i.e. minus the radial vorticity) by the flow itself. We assumed that this mode of transport can be described by a diffusion process, characterised by an effective turbulent diffusion coefficient -- or turbulent viscosity -- $\nut$, so that
\begin{equation}
    \langle \mathbf{u} ~ \Delta \Psi \rangle = -\nut \bm{\nabla} \left\langle \Delta\Psi \right\rangle_h~.
\end{equation}
This is analogous to the \citet{boussinesq1877} approximation for the Reynolds stress tensor, which is customarily extended to other moments of the form $\langle  \mathbf{u} X \rangle$ in Reynolds-averaged Navier-Stokes (RANS) models \citep[see for instance][]{xiong89a}.

Then, Eq. \ref{eqapp:StreamFunctionBeforeNut} becomes
\begin{multline}
    \left( \dfrac{\partial}{\partial t} + U\dfrac{\partial}{\partial x}\right) \Delta\Psi + \left(\beta - U''\right) \dfrac{\partial\Psi}{\partial x} - \bm{\nabla} \cdot (\nut \bm{\nabla} \Delta \langle \Psi \rangle_h) 
    \\
    + \delta \left( \dfrac{\partial \Psi}{\partial y} \dfrac{\partial \Delta\Psi}{\partial x} - \dfrac{\partial \Psi}{\partial x} \dfrac{\partial \Delta\Psi}{\partial y} \right) = 0~ .
    \label{eqapp:StreamFunctionAfterNut}
\end{multline}

\subsection{Linear inhomogeneous wave equation}

In order to derive a linear wave equation from Eq. \ref{eqapp:StreamFunctionAfterNut}, we decomposed the total stream function, $\Psi$, into the contribution from the oscillations, $\po$, and a contribution from the convective noise (i.e. the turbulence), $\pt$:
\begin{equation}
    \Psi = \po + \pt ~.
\end{equation}
In the region of excitation, we assume that
\begin{equation}
    \po \ll \pt~.
\end{equation}
This is justified in the bulk of the convective region, where small-scale convection dominates the dynamics of the star. This remains true close to the surface of the star or the tachocline: for example, in the Sun, simulations show that the typical turbulent velocities near the surface of the Sun are of the order of a few km\ s$^{-1}$ \citep{stein98}, while the typical amplitudes of  inertial modes are of the order of several m\ s$^{-1}$ at  most. However, there must be a smooth transition between the region of wave excitation and the neighbouring radiative zone, where $\pt$ is negligible. Therefore, the ordering $\po \ll \pt$ cannot be valid everywhere. In the following, we work in the region of wave excitation. We also note that, by definition of the horizontal average used to introduce the turbulent viscosity, we have

\begin{equation}
    \langle \Psi \rangle_h = \po~.
\end{equation}

Keeping only the first-order contributions in $\po$, but all orders in $\pt$, we obtain, in the region of wave excitation,
\begin{multline}
    \left( \dfrac{\partial}{\partial t} + U\dfrac{\partial}{\partial x}\right) \Delta\po + (\beta - U'') \dfrac{\partial \po}{\partial x} - \nut \Delta^2 \po \\
    + \delta\left( \dfrac{\partial \po}{\partial y} \dfrac{\partial \Delta\pt}{\partial x} + \dfrac{\partial \pt}{\partial y} \dfrac{\partial \Delta\po}{\partial x} \right. \\
    \left. - \dfrac{\partial \po}{\partial x} \dfrac{\partial \Delta\pt}{\partial y} - \dfrac{\partial \pt}{\partial x} \dfrac{\partial \Delta\po}{\partial y} \right) \\
    = -U\dfrac{\partial \Delta\pt}{\partial x} - (\beta - U'')\dfrac{\partial \pt}{\partial x} \\
    + \delta\left( \dfrac{\partial\pt}{\partial x}\dfrac{\partial \Delta\pt}{\partial y} - \dfrac{\partial\pt}{\partial y}\dfrac{\partial \Delta\pt}{\partial x} \right)~,
    \label{eqapp:FullWaveEquation}
\end{multline}
where we have gathered all inhomogeneous terms on the right-hand side: these constitute the random forcing terms. We note that we also considered $\nut$ to be uniform so that it can be pulled out of the gradient operator. The left-hand side of Eq. \ref{eqapp:WaveEquation} is split two ways: everything outside the brackets represents the deterministic linear operator governing the propagation of the waves, and everything inside the brackets represent the random fluctuations of the medium in which the waves propagate, and constitute a stochastic perturbation to the linear propagation operator.

In this study, we are interested in the excitation of the vorticity waves by turbulence, and therefore will not concern ourselves with the stochastic perturbation to the propagation operator. For this reason, we cast aside the bracket term on the left-hand side of Eq. \ref{eqapp:FullWaveEquation}. Furthermore, as is usually done while dealing with $p$-modes \citep[e.g.][]{samadi01}, we consider from the start that the linear forcing has a negligible effect on the inertial modes, because the frequencies and wavevectors in which the turbulence has significant power are far removed from those of the oscillations. This amounts to neglecting the first two terms on the right-hand side of Eq. \ref{eqapp:FullWaveEquation} (which are linear in $\pt$) compared to the bracket term (which is quadratic in $\pt$). The vorticity wave equation then becomes
\begin{multline}
    \left( \dfrac{\partial}{\partial t} + U\dfrac{\partial}{\partial x}\right) \Delta\po + (\beta - U'') \dfrac{\partial \po}{\partial x} - \nut \Delta^2 \po \\
    = \delta\left( \dfrac{\partial\pt}{\partial x}\dfrac{\partial \Delta\pt}{\partial y} - \dfrac{\partial\pt}{\partial y}\dfrac{\partial \Delta\pt}{\partial x} \right)~.
    \label{eqapp:WaveEquation}
\end{multline}

%%% Appendix B

\section{Source correlations in the frequency-wavenumber domain\label{app:SourceCorrelations}}

The contribution of the inertial modes to our synthetic power spectrum model involves the following integral (see Eq. \ref{eq:DefinitionFastVariableIntegral})
\begin{equation}
    \mathcal{I}(y_s) \equiv \displaystyle\int \d Y ~ \left\langle \widehat{S}(y_s) ~ \widehat{S}^\ast(y_s + Y) \right\rangle~,
    \label{eqapp:IntegralToCompute}
\end{equation}
where we recall that $\widehat{\vphantom{f}~.~}$ denotes the Fourier transform in $(t, x)$ and is defined by Eq. \ref{eq:DefinitionPartialFourier}, and the source term $S(t, x, y)$ is defined by the right-hand side of Eq. \ref{eq:WaveEquation}. To shorten notations, we note that the latter can be rewritten as
\begin{equation}
    S(t, x, y) = \delta\left( \epsilon_{ijz} \partial_i \pt \partial_j \Delta \pt \right)~,
    \label{eqapp:SourceTermRealSpace}
\end{equation}
where we have adopted Einstein's convention on repeated indices. Subtracting the horizontal average from the source term has, in fact, no effect on its autocorrelation spectrum, as is easily seen if the source term given by Eq. \ref{eqapp:SourceTermRealSpace} is explicitly expanded according to Eq. \ref{eqapp:DefinitionDelta}; therefore, in the following, we drop the notation $\delta$ altogether. Eq. \ref{eqapp:IntegralToCompute} then becomes
\begin{multline}
    \mathcal{I}(y_s) = \dfrac{1}{T_\mathrm{obs} X_\mathrm{obs}} \displaystyle\int \d Y \d t \d t' \d x \d x' ~ \epsilon_{ijz} \epsilon_{klz} \\
    \left\langle \left. \partial_i \pt \partial_j \Delta \pt \right|_{t, x, y_s} ~ \left. \partial_k \pt \partial_l \Delta \pt \right|_{t', x', y_s + Y} \right\rangle e^{\ii \left( \vphantom{a^a} \omega (t-t') - k_x (x-x') \right)} ~.
    \label{eqapp:IntegralToCompute2}
\end{multline}

Since the source is a quadratic function of the turbulent fluctuations of the stream function, the power spectrum ends up depending on fourth-order correlation products thereof. In the following, we make the assumption that $\pt$ and its derivatives follow a multivariate normal distribution \citep[][]{millionschchikov41}, in which case the fourth-order correlation product can be expanded in terms of second-order products only according to 
\begin{equation}
    \left\langle abcd \right\rangle = \left\langle ab \right\rangle \left\langle cd \right\rangle + \left\langle ac \right\rangle \left\langle bd \right\rangle + \left\langle ad \right\rangle \left\langle bc \right\rangle~.
\end{equation}
Then Eq. \ref{eqapp:IntegralToCompute2} becomes
\begin{equation}
    \mathcal{I}(y_s) = \mathcal{I}_a(y_s) + \mathcal{I}_b(y_s) + \mathcal{I}_c(y_s)~,
    \label{eqapp:IntegralToCompute3}
\end{equation}
where
\begin{multline}
    \mathcal{I}_a(y_s) \equiv \dfrac{1}{T_\mathrm{obs} X_\mathrm{obs}} \displaystyle\int \d Y \d t \d t' \d x \d x' ~ \epsilon_{ijz} \epsilon_{klz} \\
    \left\langle \left. \partial_i \pt \partial_j \Delta \pt \right|_{t, x, y_s} \right\rangle ~ \left\langle \left. \partial_k \pt \partial_l \Delta \pt \right|_{t', x', y_s + Y} \right\rangle \\
    e^{\ii \left( \vphantom{a^a} \omega (t-t') - k_x (x-x') \right)}~,
\end{multline}
\begin{multline}
    \mathcal{I}_b(y_s) \equiv \dfrac{1}{T_\mathrm{obs} X_\mathrm{obs}} \displaystyle\int \d Y \d t \d t' \d x \d x' ~ \epsilon_{ijz} \epsilon_{klz} \\
    \left\langle \left. \partial_i \pt \right|_{t, x, y_s} \left. \partial_k \pt \right|_{t', x', y_s + Y} \right\rangle ~ \left\langle \left. \partial_j \Delta \pt \right|_{t, x, y_s} \left. \partial_l \Delta \pt \right|_{t', x', y_s + Y} \right\rangle \\
    e^{\ii \left( \vphantom{a^a} \omega (t-t') - k_x (x-x') \right)}~,
\end{multline}
and
\begin{multline}
    \mathcal{I}_c(y_s) \equiv \dfrac{1}{T_\mathrm{obs} X_\mathrm{obs}} \displaystyle\int \d Y \d t \d t' \d x \d x' ~ \epsilon_{ijz} \epsilon_{klz} \\
    \left\langle \left. \partial_i \pt \right|_{t, x, y_s} \left. \partial_l \Delta \pt \right|_{t', x', y_s + Y} \right\rangle ~ \left\langle \left. \partial_j \Delta \pt \right|_{t, x, y_s} \left. \partial_k \pt \right|_{t', x', y_s + Y} \right\rangle \\
    e^{\ii \left( \vphantom{a^a} \omega (t-t') - k_x (x-x') \right)}~.
\end{multline}
The first integral $\mathcal{I}_a$ vanishes, because it only involves one-point, one-time correlation products, and therefore none of them depends on the time increment $t-t'$ or the space increment $x-x'$. This leaves us with only the last two integrals to consider.

First, we performed the following change of variables,
\begin{equation}
    T \equiv t~, \qquad \tau \equiv t' - t ~, \qquad X \equiv x ~, \qquad \delta x \equiv x' - x~,
\end{equation}
so that
\begin{multline}
    \mathcal{I}_b(y_s) = \dfrac{1}{T_\mathrm{obs} X_\mathrm{obs}} \displaystyle\int \d Y \d T \d \tau \d X \d \delta x ~ \epsilon_{ijz} \epsilon_{klz} \\
    \left\langle \left. \partial_i \pt \right|_{T, X, y_s} \left. \partial_k \pt \right|_{T+\tau, X+\delta x, y_s+Y} \right\rangle \\
    \left\langle \left. \partial_j \Delta \pt \right|_{T, X, y_s} \left. \partial_l \Delta \pt \right|_{T+\tau, X+\delta x, y_s+Y} \right\rangle ~ e^{\ii ( k_x \delta x - \omega \tau)}~,
\end{multline}
\begin{multline}
    \mathcal{I}_c(y_s) = \dfrac{1}{T_\mathrm{obs} X_\mathrm{obs}} \displaystyle\int \d Y \d T \d \tau \d X \d \delta x ~ \epsilon_{ijz} \epsilon_{klz} \\
    \left\langle \left. \partial_i \pt \right|_{T, X, y_s} \left. \partial_l \Delta \pt \right|_{T+\tau, X+\delta x, y_s+Y} \right\rangle \\
    \left\langle \left. \partial_j \Delta \pt \right|_{T, X, y_s} \left. \partial_k \pt \right|_{T+\tau, X+\delta x, y_s+Y} \right\rangle ~ e^{\ii ( k_x \delta x - \omega \tau)} ~.
\end{multline}
Then we consider that the two-point, two-time correlation products only depend on the time and space differences (i.e. $\tau$, $\delta x$, and $Y$), and not on the absolute time and space coordinates (i.e. $T$, $X$, and $y_s$), in which case these two integrals simplify to
\begin{multline}
    \mathcal{I}_b = \displaystyle\int \d \tau \d \delta x \d Y ~ \epsilon_{ijz} \epsilon_{klz} \left\langle \left. \partial_i \pt \right|_{0, 0, 0} \left. \partial_k \pt \right|_{\tau, \delta x, Y} \right\rangle \\
    \left\langle \left. \partial_j \Delta \pt \right|_{0, 0, 0} \left. \partial_l \Delta \pt \right|_{\tau, \delta x, Y} \right\rangle ~ e^{\ii ( k_x \delta x - \omega \tau)}
\end{multline}
and
\begin{multline}
    \mathcal{I}_c = \displaystyle\int \d \tau \d \delta x \d Y ~ \epsilon_{ijz} \epsilon_{klz} \left\langle \left. \partial_i \pt \right|_{0, 0, 0} \left. \partial_l \Delta \pt \right|_{\tau, \delta x, Y} \right\rangle \\
    \left\langle \left. \partial_j \Delta \pt \right|_{0, 0, 0} \left. \partial_k \pt \right|_{\tau, \delta x, Y} \right\rangle ~ e^{\ii ( k_x \delta x - \omega \tau)}~.
\end{multline}
If we introduce the following functions,
\begin{align}
    & f_{b, ik}(\tau, \delta\mathbf{x}) \equiv \left\langle \left. \partial_i \pt \right|_{0, \mathbf{0}} \left. \partial_k \pt \right|_{\tau, \delta \mathbf{x}} \right\rangle~, \nonumber \\
    & g_{b, jl}(\tau, \delta\mathbf{x}) \equiv \left\langle \left. \partial_j \Delta \pt \right|_{0, \mathbf{0}} \left. \partial_l \Delta \pt \right|_{\tau, \delta \mathbf{x}} \right\rangle~, \nonumber \\
    & f_{c, il}(\tau, \delta\mathbf{x}) \equiv \left\langle \left. \partial_i \pt \right|_{0, \mathbf{0}} \left. \partial_l \Delta \pt \right|_{\tau, \delta \mathbf{x}} \right\rangle~, \nonumber \\
    & g_{c, jk}(\tau, \delta\mathbf{x}) \equiv \left\langle \left. \partial_j \Delta \pt \right|_{0, \mathbf{0}} \left. \partial_k \pt \right|_{\tau, \delta \mathbf{x}} \right\rangle~,
    \label{eqapp:DefinitionFunctionsFG}
\end{align}
where $\delta \mathbf{x} \equiv \delta x \mathbf{e}_x + Y \mathbf{e}_y$, then these two integrals can be rewritten more compactly as
\begin{align}
    & \mathcal{I}_b = \sqrt{T_\mathrm{obs} X_\mathrm{obs} Y_\mathrm{obs}} \epsilon_{ijz} \epsilon_{klz} \widetilde{f_{b, ik} g_{b, jl}}(-\omega, -k_x, 0)~, \\
    & \mathcal{I}_c = \sqrt{T_\mathrm{obs} X_\mathrm{obs} Y_\mathrm{obs}} \epsilon_{ijz} \epsilon_{klz} \widetilde{f_{c, il} g_{c, jk}}(-\omega, -k_x, 0)~,
\end{align}
where the notation $\widetilde{\vphantom{f} ~.~}$ denotes the Fourier transform in $(t, x, y)$, and is defined by
\begin{equation}
    \widetilde{f}(\omega, k_x, k_y) \equiv \dfrac{1}{\sqrt{T_\mathrm{obs} X_\mathrm{obs} Y_\mathrm{obs}}} \displaystyle\int \d t \d x \d y ~ f(t, x, y) e^{\ii(\omega t - k_x x - k_y y)}~. \label{eqapp:DefinitionFullFourier}
\end{equation}
Then, expanding the Fourier transform of the products as convolution integrals, and exploiting the fact that $\widetilde{g}(-\omega, -\mathbf{k}) = \widetilde{g}^\ast(\omega, \mathbf{k})$ we obtain
\begin{multline}
    \mathcal{I}_b = \dfrac{T_\mathrm{obs} X_\mathrm{obs} Y_\mathrm{obs}}{(2\pi)^3} \displaystyle\int \d\omega' \d^2\mathbf{k}' \epsilon_{ijz} \epsilon_{klz} \widetilde{f_{b, ik}}(\omega', \mathbf{k}') \\
    \widetilde{g_{b, jl}}^\ast(\omega' + \omega, \mathbf{k}' + \mathbf{k})~,
    \label{eqapp:Ib}
\end{multline}
\begin{multline}
    \mathcal{I}_c = \dfrac{T_\mathrm{obs} X_\mathrm{obs} Y_\mathrm{obs}}{(2\pi)^3} \displaystyle\int \d\omega' \d^2\mathbf{k}' \epsilon_{ijz} \epsilon_{klz} \widetilde{f_{c, il}}(\omega', \mathbf{k}') \\
    \widetilde{g_{c, jk}}^\ast(\omega' + \omega, \mathbf{k}' + \mathbf{k})~,
    \label{eqapp:Ic}
\end{multline}
where $\mathbf{k} \equiv k_x \mathbf{e_x}$.

Next, we need to express the quantities $\widetilde{f_{b, ik}}$, $\widetilde{g_{b, jl}}$, $\widetilde{f_{c, il}}$ and $\widetilde{g_{c, jk}}$ appearing in these integrals. Taking the Fourier transforms of each of Eq. \ref{eqapp:DefinitionFunctionsFG}, we find
\begin{multline}
    \widetilde{f_{b, ik}}(\omega', \mathbf{k}') = \dfrac{\sqrt{T_\mathrm{obs} X_\mathrm{obs} Y_\mathrm{obs}}}{(2\pi)^3} \displaystyle\int \d\omega'' \d^2\mathbf{k}'' ~ (-k_i'' k_k') \\
    \left\langle \ffpt(\omega'', \mathbf{k}'') ~ \ffpt(\omega', \mathbf{k}') \right\rangle~,
\end{multline}
\begin{multline}
    \widetilde{g_{b, jl}}(\omega', \mathbf{k}') = \dfrac{\sqrt{T_\mathrm{obs} X_\mathrm{obs} Y_\mathrm{obs}}}{(2\pi)^3} \displaystyle\int \d\omega'' \d^2\mathbf{k}'' ~ (-k_j'' k''^2 k_l' k'^2) \\
    \left\langle \ffpt(\omega'', \mathbf{k}'') ~ \ffpt(\omega', \mathbf{k}') \right\rangle~,
\end{multline}
\begin{multline}
    \widetilde{f_{c, il}}(\omega', \mathbf{k}') = \dfrac{\sqrt{T_\mathrm{obs} X_\mathrm{obs} Y_\mathrm{obs}}}{(2\pi)^3} \displaystyle\int \d\omega'' \d^2\mathbf{k}'' ~ (+k_i'' k_l' k'^2) \\
    \left\langle \ffpt(\omega'', \mathbf{k}'') ~ \ffpt(\omega', \mathbf{k}') \right\rangle~,
\end{multline}
and
\begin{multline}
    \widetilde{g_{c, jk}}(\omega', \mathbf{k}') = \dfrac{\sqrt{T_\mathrm{obs} X_\mathrm{obs} Y_\mathrm{obs}}}{(2\pi)^3} \displaystyle\int \d\omega'' \d^2\mathbf{k}'' ~ (+k_j'' k''^2 k_k') \\
    \left\langle \ffpt(\omega'', \mathbf{k}'') ~ \ffpt(\omega', \mathbf{k}') \right\rangle~.
\end{multline}
Because we have assumed a homogeneous and stationary turbulence, the correlation products can be rewritten in terms of Dirac distributions, yielding non-zero contributions only if $\omega'' = -\omega'$ and $\mathbf{k}'' = \mathbf{k}'$. We obtain
\begin{align}
    & \widetilde{f_{b, ik}}(\omega', \mathbf{k}') = \dfrac{1}{\sqrt{T_\mathrm{obs} X_\mathrm{obs} Y_\mathrm{obs}}} k_i' k_k' \ep(\omega', \mathbf{k}')~, \\
    & \widetilde{g_{b, jl}}(\omega', \mathbf{k}') = \dfrac{1}{\sqrt{T_\mathrm{obs} X_\mathrm{obs} Y_\mathrm{obs}}} k_j' k_l' k'^4 \ep(\omega', \mathbf{k}')~, \\
    & \widetilde{f_{c, il}}(\omega', \mathbf{k}') = -\dfrac{1}{\sqrt{T_\mathrm{obs} X_\mathrm{obs} Y_\mathrm{obs}}} k_i' k_l' k'^2 \ep(\omega', \mathbf{k}')~, \\
    & \widetilde{g_{c, jk}}(\omega', \mathbf{k}') = -\dfrac{1}{\sqrt{T_\mathrm{obs} X_\mathrm{obs} Y_\mathrm{obs}}} k_j' k_k' k'^2 \ep(\omega', \mathbf{k}')~.
\end{align}
The function $\ep$ denotes the stream function turbulent spectrum, defined by
\begin{equation}
    \ep(\omega', \mathbf{k}') \equiv \displaystyle\int \d\tau \d^2\mathbf{x} \left\langle \pt(T, \mathbf{X}) \pt(T+\tau, \mathbf{X}+\mathbf{x}) \right\rangle ~ e^{\ii (\omega'\tau - \mathbf{k}' \cdot \mathbf{x})}~.
    \label{eqapp:DefinitionStreamFunctionSpectrum}
\end{equation}
Plugging these into Eq. \ref{eqapp:Ib} and Eq. \ref{eqapp:Ic}, we find
\begin{multline}
    \mathcal{I}_b = \dfrac{1}{(2\pi)^3} \displaystyle\int \d\omega' \d^2\mathbf{k}' \epsilon_{ijz} \epsilon_{klz} k_i' (k_j'+k_j) k_k' (k_l'+k_l) \\
    (\mathbf{k}' + \mathbf{k})^4 ~ \ep(\omega', \mathbf{k}') \ep^\ast(\omega' + \omega, \mathbf{k}' + \mathbf{k})~,
\end{multline}
and
\begin{multline}
    \mathcal{I}_c = \dfrac{1}{(2\pi)^3} \displaystyle\int \d\omega' \d^2\mathbf{k}' \epsilon_{ijz} \epsilon_{klz} k_i' (k_j' + k_j)(k_k' + k_k) k_l' k'^2 \\
    (\mathbf{k}' + \mathbf{k})^2 \ep(\omega', \mathbf{k}') \ep^\ast(\omega' + \omega, \mathbf{k}' + \mathbf{k})~.
\end{multline}
Flipping the indices $k$ and $l$ in the second integral, and exploiting the fact that $\epsilon_{klz} = -\epsilon_{lkz}$, the sum of these two integrals becomes
\begin{multline}
    \mathcal{I} = \dfrac{1}{(2\pi)^3} \displaystyle\int \d\omega' \d^2\mathbf{k}' \epsilon_{ijz} \epsilon_{klz} k_i' (k_j'+k_j) k_k' (k_l'+k_l) \\
    (\mathbf{k}' + \mathbf{k})^2 \left( (\mathbf{k}' + \mathbf{k})^2 - \mathbf{k}'^2 \right) ~ \ep(\omega', \mathbf{k}') \ep^\ast(\omega' + \omega, \mathbf{k}' + \mathbf{k})~.
\end{multline}

Finally, we explicitly expanded the index contractions. A basic property of the Levi-Civita symbol is
\begin{equation}
    \epsilon_{ijz} \epsilon_{klz} = \delta_{ik}\delta_{jl}\delta_{zz} + \delta_{il}\delta_{jz}\delta_{kz} + \delta_{iz}\delta_{jk}\delta_{lz} - \delta_{ik}\delta_{jz}\delta_{lz} - \delta_{iz}\delta_{jl}\delta_{kz} - \delta_{il}\delta_{jk}\delta_{zz}~.
    \label{eqapp:IndexContraction}
\end{equation}
Seeing as neither $\mathbf{k}'$ nor $\mathbf{k}$ has a component along the $z$-axis, we have
\begin{align}
    \epsilon_{ijz} \epsilon_{klz} k_i' (k_j' + k_j) k_k' (k_l' + k_l) & = (\delta_{ik}\delta_{jl} - \delta_{il}\delta_{jk}) k_i' (k_j' + k_j) k_k' (k_l' + k_l) \nonumber \\
    & = k'^2 (\mathbf{k}' + \mathbf{k})^2 - (\mathbf{k}'\cdot(\mathbf{k}' + \mathbf{k}))^2 \nonumber \\
    & = k'^2 k^2 - (\mathbf{k}' \cdot \mathbf{k})^2~.
\end{align}
Thus we obtain
\begin{multline}
    \mathcal{I} = \dfrac{1}{(2\pi)^3} \displaystyle\int \d\omega' \d^2\mathbf{k}' ~ \left[ \vphantom{a^a} k'^2 k^2 - (\mathbf{k}' \cdot \mathbf{k})^2 \right] (\mathbf{k}' + \mathbf{k})^2 \\
    \left( (\mathbf{k}' + \mathbf{k})^2 - k'^2 \right) ~ \ep(\omega', \mathbf{k}') \ep^\ast(\omega' + \omega, \mathbf{k}' + \mathbf{k})~.
\end{multline}
This expression can be rendered more symmetric by shifting the two variables $\omega'$ and $\mathbf{k}'$ by $\omega/2$ and $\mathbf{k}/2$, respectively, which leads to the final expression for the integral $\mathcal{I}$ as a function of the stream function turbulent spectrum, $\ep$,
\begin{multline}
    \mathcal{I} = \dfrac{1}{4\pi^3} \displaystyle\int \d\omega' \d^2\mathbf{k}' ~ k_x^3 k_x' k_y'^2 \left| \mathbf{k}' + \dfrac{\mathbf{k}}{2} \right|^2 \\
    \ep\left(\omega' - \dfrac{\omega}{2}, \mathbf{k}' - \dfrac{\mathbf{k}}{2} \right) \ep^\ast\left(\omega' + \dfrac{\omega}{2}, \mathbf{k}' + \dfrac{\mathbf{k}}{2} \right)~.
    \label{eqapp:FastVariableIntegral}
\end{multline}

%%% Appendix C

\section{Chebyshev representation of the linear operator\label{app:LinearMatrixGreen}}

In this Appendix, we derive the components of the linear operator $\mathcal{L}$, defined by Eq. \ref{eq:DefinitionLinearOperator}, in the dual basis formed by the Chebyshev polynomials of the first kind. The derivation follows the work by \citet{orszag71}. We recall that the Chebyshev polynomials of the first kind are defined by
\begin{equation}
    T_n(\xi) = \cos (n \arccos \xi)~.
    \label{eqapp:DefinitionChebyshev}
\end{equation}
Using the definition of the linear operator, we can write
\begin{align}
    \mathcal{M}_{ij} =& - \left(\omega k_x^2 + k_x(\beta - U'') + \ii \nut k_x^4 \right) \mathcal{M}_{ij}^{(1)} \\
    &+ \left(\omega + 2 \ii \nut k_x^2\right) \mathcal{M}_{ij}^{(2)} \\
    &- \ii \nut \mathcal{M}_{ij}^{(3)} + k_x^3 \mathcal{M}_{ij}^{(4)} - k_x \mathcal{M}_{ij}^{(5)}~,
\end{align}
where
\begin{align}
    & \mathcal{M}_{ij}^{(1)} \equiv \dfrac{2}{\pi c_i} \bigbraket{T_i}{T_j} = \delta_{ij}~, \\
    & \mathcal{M}_{ij}^{(2)} \equiv \dfrac{2}{\pi c_i} \bigbraket{T_i}{T_j''}~, \\
    & \mathcal{M}_{ij}^{(3)} \equiv \dfrac{2}{\pi c_i} \bigbraket{T_i}{T_j''''}~, \\
    & \mathcal{M}_{ij}^{(4)} \equiv \dfrac{2}{\pi c_i} \bigbraket{T_i}{UT_j}~, \\
    & \mathcal{M}_{ij}^{(5)} \equiv \dfrac{2}{\pi c_i} \bigbraket{T_i}{U T_j''}~,
\end{align}
and $T_j''$ and $T_j''''$ denote the second and fourth derivatives with respect to $\xi$, respectively. We derive $\mathcal{M}_{ij}^{(2)}$ and $\mathcal{M}_{ij}^{(3)}$ in Sect. \ref{subapp:ProjectionDerivatives}, and $\mathcal{M}_{ij}^{(4)}$ and $\mathcal{M}_{ij}^{(5)}$ in Sect. \ref{subapp:ProjectionProducts}.

\subsection{Projection of the derivatives\label{subapp:ProjectionDerivatives}}
 
Given a function
\begin{equation}
    f (\xi) = \sum \lambda_i T_i (\xi)~,
\end{equation}
we can write, for any positive integer, $p$,
\begin{equation}
    \dfrac{\d^p f}{\d y^p} = \sum_{i=0}^{+\infty} \lambda_i^{(p)} T_i~, \qquad \lambda_i^{(0)} = \lambda_i~.
 \end{equation}
In order to compute the coefficients $\mathcal{M}_{ij}^{(2)}$ and $\mathcal{M}_{ij}^{(3)}$, we needed to find the recurrence relation between the $\lambda_i^{(p)}$ and the $\lambda_i^{(p-1)}$, that is, a recurrence relation for all the derivatives of $f$.

For any $p \geqslant 1$ we can write
\begin{equation}
    \sum_{i=0}^{+\infty} \lambda_i^{(p)} T_i = \dfrac{\d}{\d y} \sum_{i=0}^{+\infty} \lambda_i^{(p-1)} T_i~.
    \label{eqapp:TempRecurrence1}
\end{equation}
On the other hand, from Eq. \ref{eqapp:DefinitionChebyshev}, it is easily seen that
\begin{equation}
    T_i = \dfrac{c_i}{2(i+1)}T_{i+1}' - \dfrac{c_i'}{2(i-1)}T_{i-1}'~,
\end{equation}
where we recall that $c_i = 1 + \delta_{i0}$, and we have also introduced a new factor $c_i'$ (defined by $c_0' = c_1' = 0$ and $c_{i \ge 2}' = 1$). Therefore, we can also write
\begin{align}
    \sum_{i=0}^{+\infty} \lambda_i^{(p)} T_i 
    & = \dfrac{\d}{\d y} \sum_{i=0}^{+\infty} \lambda_i^{(p)} \left( \dfrac{c_i}{2(i+1)} T_{i+1} - \dfrac{c_i'}{2(i-1)}T_{i-1} \right) \nonumber \\
    & = \dfrac{\d}{\d y} \sum_{i=1}^{+\infty} \left( \dfrac{c_{i-1}}{2i} \lambda_{i-1}^{(p)} - \dfrac{1}{2i} \lambda_{i+1}^{(p)} \right) T_i~.
    \label{eqapp:TempRecurrence2}
\end{align}
Equating the $T_i$ coefficients in Eqs. \ref{eqapp:TempRecurrence1} and \ref{eqapp:TempRecurrence2}, we find
\begin{equation}
    c_{i-1} \lambda_{i-1}^{(p)} - \lambda_{i+1}^{(p)} = 2i\lambda_i^{(p-1)}~,
\end{equation}
which constitutes a recurrence relation in terms of $p$. This recurrence relation, combined with the condition that we must have $\lim_{i \to +\infty} \lambda_i^{(p)} = 0$ for any value of $p$, is easily solved to yield
\begin{equation}
    c_i \lambda_i^{(p)} = 2 \sum_{\substack{j=i+1 \\ i+j \equiv 1[2]}}^{+\infty} j \lambda_j^{(p-1)}~.
\end{equation}
where the notation $\equiv a [b]$ means `equal to $a$ modulo $b$'.

For example, the coefficients of the first derivative straightforwardly read
\begin{equation}
    c_i \lambda_i^{(1)} = 2 \sum_{\substack{j=i+1 \\ i+j \equiv 1[2]}}^{+\infty} j \lambda_j~.
\end{equation}
The coefficients of the second derivative can be computed in the following way
\begin{align}
    c_i \lambda_i^{(2)}
    & = 2 \sum_{\substack{j=i+1 \\ i+j \equiv 1[2]}}^{+\infty} j \lambda_j^{(1)} = 4 \sum_{\substack{j=i+1 \\ i+j \equiv 1[2]}}^{+\infty} j \sum_{\substack{k=j+1 \\ k+j \equiv 1 [2]}}^{+\infty} k\lambda_k \nonumber \\
    & = 4\sum_{\substack{k=i+2 \\ k+i \equiv 0 [2]}}^{+\infty} k \lambda_k \sum_{\substack{j=i+1 \\ j+k \equiv 1 [2]}}^{k-1} j = \sum_{\substack{k=i+2 \\ k+i \equiv 0 [2]}}^{+\infty} k \left( k^2 - i^2 \right) \lambda_k~.
\end{align}
Similarly, we find
\begin{align}
    & c_i \lambda_i^{(3)} = \dfrac{1}{4} \sum_{\substack{k=i+3 \\ k+i \equiv 1 [2]}}^{+\infty} k \left[ k^2 \left( k^2 - 2 \right) - 2 k^2 i^2 + i^2 \left( i^2 - 2 \right) + 1 \right] \lambda_k~, \\
    & c_i \lambda_i^{(4)} = \dfrac{1}{24} \sum_{\substack{k=i+4 \\ k+i \equiv 0 [2]}}^{+\infty} k \left[ k^2 \left( k^2 - 4 \right)^2 - 3 k^4 i^2 + 3 k^2 i^4 - i^2 \left( i^2 - 4 \right)^2 \right] \lambda_k~.
\end{align}
From these expressions it directly follows that
\begin{equation}
    \mathcal{M}_{ij}^{(2)} =
    \begin{cases}
        \dfrac{1}{c_i} j \left( j^2 - i^2 \right) & \mathrm{if} ~~ j \geqslant i+2 ~ \mathrm{and} ~ i+j \equiv 0 ~ [2] \\
        0 & \mathrm{otherwise}~,
    \end{cases}
\end{equation}
and
\begin{equation}
    \mathcal{M}_{ij}^{(3)} =
    \begin{cases}
        \dfrac{1}{24 c_i} j \left[ j^2 \left( j^2 - 4 \right)^2 - 3 j^4 i^2 + 3 j^2 i^4 - i^2 \left( i^2 - 4 \right)^2 \right] \\
        \hspace{2cm} \mathrm{if} ~~ j \geqslant i+4 ~ \mathrm{and} ~ i+j \equiv 0 ~ [2] & \\
        0 \hspace{1.85cm} \mathrm{otherwise} & ~.
    \end{cases}
\end{equation}

\subsection{Projection of the products\label{subapp:ProjectionProducts}}

In order to compute the coefficients $\mathcal{M}_{ij}^{(4)}$ and $\mathcal{M}_{ij}^{(5)}$, we need a rule for the expansion of a product on the Chebyshev basis. We now show that this rule takes the form of a convolution. We considered two arbitrary functions $f$ and $g$, expanded on the Chebyshev polynomials according to
\begin{align}
    & f(\xi) = \sum_{i=0}^{+\infty} \lambda_i T_i(\xi)~, \\
    & g(\xi) = \sum_{i=0}^{+\infty} \mu_i T_i(\xi)~.
\end{align}
It will be useful, in the following, to introduce the following functions
\begin{equation}
    \widetilde{T}_n(\xi) \equiv e^{\ii n \arccos\xi}~,
\end{equation}
so that we have
\begin{equation}
    2T_i = \widetilde{T}_i + \widetilde{T}_{-i}~.
    \label{eqapp:IdentityTtildeT}
\end{equation}
It directly follows that
\begin{align}
    & 2f = \sum_{i=-\infty}^{+\infty} c_{|i|} \lambda_{|i|} \widetilde{T}_i~, \\
    & 2g = \sum_{i=-\infty}^{+\infty} c_{|i|} \mu_{|i|} \widetilde{T}_i~.
\end{align}
Forming the product of these two expressions, and using the identity $\widetilde{T}_i\widetilde{T}_j = \widetilde{T}_{i+j}$, we obtain
\begin{equation}
    4fg = \sum_{i=-\infty}^{+\infty} \sum_{j=-\infty}^{+\infty} c_{|i|} c_{|j|} \lambda_{|i|} \mu_{|j|} \widetilde{T}_{i+j}~,
\end{equation}
which can be rewritten as
\begin{equation}
    4fg = \sum_{i=-\infty}^{+\infty} \left( \sum_{j=-\infty}^{+\infty} c_{|j|} c_{|i-j|} \lambda_{|j|} \mu_{|i-j|} \right) \widetilde{T}_i~.
\end{equation}
Finally, using Eq. \ref{eqapp:IdentityTtildeT} to revert back to an expansion on the $T_i$, one finally finds
\begin{equation}
    fg = \sum_{i=0}^{+\infty} \left( \dfrac{1}{2c_i} \sum_{j=-\infty}^{+\infty} c_{|j|} c_{|i-j|} \lambda_{|j|} \mu_{|i-j|} \right) T_i~.
\end{equation}

The differential rotation profile in the expressions of $\mathcal{M}_{ij}^{(4)}$ and $\mathcal{M}_{ij}^{(5)}$ is given by $U(\xi) = -\overline{U} \xi^2$, which is straightforwardly expanded as
\begin{equation}
    U = -\dfrac{1}{2}\overline{U} \left( T_0 + T_2 \right)~, 
\end{equation}
so that
\begin{equation}
    fU = -\dfrac{\overline{U}}{4} \sum_{i=0}^{+\infty} \dfrac{1}{c_i} \left( \lambda_{i+2} + 2c_i\lambda_i + c_{|i-2|}\lambda_{|i-2|} \right) T_i~.
\end{equation}
From this expression directly stem the following
\begin{align}
    & \mathcal{M}_{ij}^{(4)} = -\dfrac{\overline{U}}{4c_i} \left( \delta_{i+2,j} + 2 c_i \delta_{ij} + c_{|i-2|} \delta_{|i-2|,j} \right)~, \\
    & \mathcal{M}_{ij}^{(5)} = -\dfrac{\overline{U}}{4c_i} \left( \mathcal{M}_{i+2,j}^{(2)} + 2 c_i \mathcal{M}_{ij}^{(2)} + c_{|i-2|} \mathcal{M}_{|i-2|,j}^{(2)} \right)~.
\end{align}

\end{appendix}

\end{document}